\documentclass{mn2e}
\pdfoutput=1
\def\eg{\hbox{\it eg. }}
\def\etal{\hbox{\it et al. }}
\def\spose#1{\hbox to 0pt{#1\hss}}
\def\approxgt{\mathrel{\spose{\lower 3pt\hbox{$\sim$}}\raise 2.0pt\hbox{$>$}}}
\def\approxlt{\mathrel{\spose{\lower 3pt\hbox{$\sim$}}\raise 2.0pt\hbox{$<$}}}
\def\apc{\rm atom cm$^{-2}$}
\def\ie{{\it i.e.\ }}
\usepackage{graphics,graphicx}

\title[Complex X-ray emission from merging clusters]
{Chandra measurements of non-thermal-like
X-ray emission from massive, merging, radio-halo clusters}
\author[E. T. Million and S. W. Allen]
{\parbox[]{6.in} {E. T. Million and S. W. Allen\\
\footnotesize
Kavli Institute for Particle Astrophysics and Cosmology, Stanford University, 382 Via Pueblo Mall, Stanford, CA 94305-4060, USA. \\
}}
\begin{document}
\renewcommand{\thefootnote}{\arabic{\footnote}}
\maketitle
\begin{abstract}
We report the discovery of spatially-extended, non-thermal-like
emission components in Chandra X-ray spectra for five of
a sample of seven massive, merging galaxy clusters with powerful radio halos. 
The emission
components can be fitted by power-law models with mean photon indices
in the range $1.5<\Gamma<2.0$.  A control sample of regular,
dynamically relaxed clusters, without radio halos but with comparable mean
thermal temperatures and luminosities, shows no compelling evidence for similar
components. Detailed X-ray spectral mapping
reveals the complex thermodynamic states of the radio halo clusters.
Our deepest observations, of the Bullet Cluster 1E\,0657-56,
demonstrate a spatial correlation between the strongest power-law
X-ray emission, highest thermal pressure, and brightest 1.34GHz radio
halo emission in this cluster.  We confirm the presence of a shock front
in the 1E\,0657-56 and report the discovery of a new, 
large-scale shock front in Abell 2219.
We explore possible origins for the power-law 
X-ray components. These include inverse Compton scattering of
cosmic microwave background photons by relativistic electrons in the
clusters; bremsstrahlung from supra-thermal electrons
energized by Coulomb collisions with an energetic, non-thermal proton
population; and synchrotron emission associated with ultra-relativistic
electrons. Interestingly, we show that the power-law 
signatures may also be due to complex temperature and/or
metallicity structure in clusters particularly in the presence of 
metallicity gradients.
In this case, an important distinguishing characteristic between the radio
halo clusters and control sample of predominantly cool-core clusters
is the relatively low central X-ray surface brightness of the former.  
Our results have implications for previous discussions of soft excess X-ray
emission from clusters and highlight
the importance of further deep X-ray and radio mapping,
coupled with new hard X-ray, $\gamma$-ray and TeV observations, for
improving our understanding of the non-thermal particle populations in
these systems.
\end{abstract}

\begin{keywords}
X-rays: galaxies: clusters -- galaxies: clusters: individual: 1E\,0657-56, 
A665, A2163, A2255, A2319, A2744, A2219, A576, A1795, A2204, 
A478, A2029 --
radiation mechanisms: non-thermal -- 
intergalactic medium -- magnetic fields 
\end{keywords}

\section{Introduction}

Combined X-ray, optical and radio studies of massive, X-ray luminous
galaxy clusters demonstrate a clear correlation between the presence
of spatially extended radio halo emission and strong, recent merger
activity (\eg Buote 2001; Schuecker \etal 2001; Govoni \etal 2001;
Giovannini \& Feretti 2002).  The observed radio halo emission is
synchrotron radiation from relativistic electrons, thought to have
been re-accelerated by turbulence and/or shocks resulting from 
merger activity (\eg Ensslin \etal 1998; Bykov \etal 2000; Brunetti
\etal 2001; Ensslin \& Br\"uggen 2002; Petrosian 2003; Brunetti \etal
2004; Brunetti \& Blasi 2005; Feretti 2005; Cassano, Brunetti \& Setti
2006; Cassano \etal 2007; Brunetti \& Lazarian 2007; Pfrommer 2008;
Pfrommer, Ensslin \& Springel 2008; Wolfe \& Melia 2008; Petrosian \&
Bykov 2008).  The surface brightness of the radio halo emission is
typically low (of order $\mu$Jy per arcsec$^2$) but can span regions
of a Mpc or more in size (\eg Feretti 2005; Feretti \etal 2005).

The same population of relativistic electrons responsible for the
radio halo emission is also expected to produce non-thermal X-ray
emission via Inverse Compton (IC) scattering of the Cosmic Microwave
Background (CMB). In principle, the combination of radio and IC X-ray
flux measurements can then be used to determine, or even map, the
magnetic field strength in merging clusters (\eg Rephaeli 1979;
Fusco-Femiano \etal 1999; Eckert \etal 2008). However, other processes
can also contribute to a non-thermal-like X-ray emission
signature,\footnote{ By non-thermal-like X-ray emission, we mean emission
over and above that expected from the diffuse, ambient cluster gas that can
be approximated by a power-law.}
including shocks (which can temporarily heat regions of massive,
merging clusters to temperatures of tens of keV e.g. Markevitch \&
Vikhlinin 2007 and references therein; see also Komatsu \etal 2001;
Allen \etal 2002), a range of non-thermal bremsstrahlung processes
(\eg Ensslin, Lieu \& Biermann 1999; Sarazin \& Kempner 2000;
Petrosian 2001; Liang \etal 2002; Dogiel \etal 2007; Wolfe \& Melia
2008) and synchrotron emission by ultra-relativistic electrons and
positrons (Timokhin, Aharonian \& Neronov 2004; Inoue, Aharonian \&
Sugiyama 2005). Distinguishing such emission components in the
presence of potentially complicated astrophysical and instrumental
backgrounds, and the luminous, diffuse, ambient thermal X-ray emission
from clusters, is challenging.

At hard X-ray wavelengths, a detection of non-thermal X-ray emission
from the nearby Coma Cluster has been reported by Fusco-Femiano \etal
(1999; see also Fusco-Femiano \etal 2004) using BeppoSAX Phoswich
Detection System (PDS) observations. The Coma cluster contains a
powerful, extended radio halo and, being one of the nearest, X-ray
brightest clusters, is an excellent candidate for the detection of IC
X-ray emission. However, complexities associated with
modelling the PDS data have led Rossetti \& Molendi 2004 (see also
Rossetti \& Molendi 2007) to question the significance of this
detection (but see also Fusco-Femiano,
Landi \& Orlandini 2007). Rephaeli, Gruber \& Blanco (1999) and
Rephaeli \& Gruber (2002) also report a detection of
non-thermal X-ray emission from the Coma cluster, based on Rossi X-ray
Timing Explorer (RXTE) data.  However, the large angular beam of the
RXTE instruments also makes it difficult to disentangle true
non-thermal emission components from spatial variations in the
spectrum of the diffuse intracluster gas.  
More recent observations of the Coma Cluster with 
$INTEGRAL$, $Swift$/BAT and Suzaku, 
argue that there is no significant non-thermal
X-ray emission at high energies in the Coma cluster
(see Renaud \etal 2006; Ajello \etal 2009; Wik \etal 2009).
Eckert \etal (2008) report
a detection of hard X-ray emission from the Ophiuchus cluster using
observations made with the $INTEGRAL$ satellite. 
However, more recent $Swift$/BAT and Suzaku
analyses of the cluster do not confirm this result (Ajello \etal 2009; 
Fujita \etal 2008).  Clearly the situation with regard to hard X-ray
emission from clusters remains uncertain and more and better data are required.

Soft X-ray (energy $E<0.5$\,keV) and far ultraviolet observations
offer another avenue to search for non-thermal X-ray emission.
However, measurements at these wavelengths require precise
background modelling and detailed knowledge of 
the thermal emission spectrum and 
Galactic column density along the line of sight (\eg
Fabian 1996). Although strong arguments have been made for the
presence of excess UV/soft X-ray emission in the Coma Cluster (\eg
Lieu \etal 1996; Bowyer, Bergh\"ofer \& Korpela 1999; Arabadjis \&
Bregman 2000; Bonamente \etal 2001a; Bonamente \etal 2003; Finoguenov
\etal 2003; Bowyer \etal 2004), for other systems, claims of diffuse
UV emission over and above that expected from the intracluster plasma,
remain controversial (see \eg Bregman \& Lloyd-Davies 2006; see
also Durret \etal 2008).

New and forthcoming $\gamma$-ray and hard X-ray missions such as the {\it Fermi
Gamma-ray Space Telescope (Fermi)} and the {\it Nuclear
Spectroscopic Telescope Array (NuSTAR)} should shed much light on the
non-thermal and quasi-thermal particle populations in galaxy
clusters. However, at present, arguably the best waveband in which to
search for non-thermal and quasi-thermal emission signatures 
in galaxy clusters remains
the $\sim 0.5-8.0$ keV energy range covered by major X-ray
observatories such as Chandra and XMM-Newton.  In particular, the
Chandra X-ray Observatory offers excellent spatial resolution, good
spectral resolution, is well calibrated and has a well understood
instrumental background. The Chandra observing band extends down to
energies soft enough to allow Galactic column densities to be measured
precisely, and is wide enough that `non-thermal' X-ray emission (be it
true power-law-like emission or complexities in the thermal emission
spectrum) can, in principle, be detected.  In particular, detailed
spatially-resolved X-ray spectroscopy with Chandra minimizes
systematic uncertainties associated with spatial variations in the
thermal spectrum of the diffuse cluster gas (although, as we show
here, does not eliminate them). Such considerations have
motivated the present study, in which we have used Chandra to carry
out detailed spatially-resolved spectral mapping of the most X-ray
luminous merging, radio halo clusters known, with the goal of
searching for quasi-thermal and/or non-thermal X-ray emission.

The structure of this paper is as follows. In Section 2, we describe
the target selection, data reduction, region selection, and spectral
analysis.  Section 3 summarizes the evidence for complex non-thermal-like X-ray
emission components in the radio-halo clusters. Section 4 reports
results from detailed thermodynamic mapping of the clusters, with
particular emphasis paid to the two deepest observations, of the
Bullet Cluster, 1E\,0657-56, and the relaxed, control cluster Abell
2029.  Section 5 discusses the possible origins of the observed complex
X-ray components. Section 6 summarizes our conclusions. A
$\Lambda$CDM cosmological model with $H_0=70$ km s$^{-1}$ Mpc$^{-1}$,
$\Omega_{\rm m}=0.3$, and $\Omega_{\Lambda}=0.7$ is assumed
throughout.

\section{X-Ray Observations and Analysis}
\label{section:obs}

\begin{table*}
\begin{center}
\caption{Summary of the Chandra observations. Columns list the target name, 
redshift, observation ID, detector, observation mode, exposure after cleaning 
and observation date.}\label{table:obs}
\vskip 0 truein
\begin{tabular}{ c c c c c c c c c }
&&&&&&&&\\
Name                 & ~ & z & & Obs. ID & Detector & Mode & Exposure (ks) & Observation Date\\
\hline \vspace{-0.2cm}\\
\multicolumn{1}{c}{} &
\multicolumn{1}{c}{} &
\multicolumn{1}{c}{} &
\multicolumn{4}{c}{\hspace{2cm} RADIO HALO CLUSTERS} &
\multicolumn{1}{c}{} &
\multicolumn{1}{c}{} \\
\vspace{-0.1cm}
&&&&&&&&\\

Abell 2319           & ~ & 0.056  &&  3231  & ACIS-I & VFAINT & 14.4 & Mar. 14 2002\\
Abell 2255           & ~ & 0.0809 &&  894   & ACIS-I & FAINT  & 38.2 & Oct. 20 2000\\
Abell 665            & ~ & 0.182  &&  3586  & ACIS-I & VFAINT & 15.9 & Dec. 28 2002\\
Abell 2163           & ~ & 0.203  &&  1653  & ACIS-I & VFAINT & 62.4 & Jun. 16 2001\\
Abell 2219           & ~ & 0.228  &&  896   & ACIS-S & FAINT  & 41.3 & Mar. 31 2000\\
1E\,0657-56 (1)      & ~ & 0.297  &&  5356  & ACIS-I & VFAINT & 96.2 & Aug. 11 2004\\
1E\,0657-56 (2)      & ~ & "      &&  3184  & ACIS-I & VFAINT & 82.4 & Jul. 12 2002\\
1E\,0657-56 (3)      & ~ & "      &&  5361  & ACIS-I & VFAINT & 82.1 & Aug. 17 2004\\
1E\,0657-56 (4)      & ~ & "      &&  5357  & ACIS-I & VFAINT & 78.3 & Aug. 14 2004\\
1E\,0657-56 (5)      & ~ & "      &&  4984  & ACIS-I & VFAINT & 75.9 & Aug. 19 2004\\
1E\,0657-56 (6)      & ~ & "      &&  4986  & ACIS-I & VFAINT & 40.7 & Aug. 25 2004\\
1E\,0657-56 (7)      & ~ & "      &&  5358  & ACIS-I & VFAINT & 31.7 & Aug. 15 2004\\
1E\,0657-56 (8)      & ~ & "      &&  4985  & ACIS-I & VFAINT & 24.0 & Aug. 23 2004\\
1E\,0657-56 (9)      & ~ & "      &&  5355  & ACIS-I & VFAINT & 22.3 & Aug. 10 2004\\
Abell 2744           & ~ & 0.308  &&  2212  & ACIS-S & VFAINT & 20.8 & Sep. 3 2001\\
Abell 2744	     & ~ & "      &&  7915  & ACIS-I & VFAINT & 14.5 & Nov. 11 2006\\
Abell 2744	     & ~ & "      &&  8477  & ACIS-I & VFAINT & 42.3 & Jun. 10 2007\\
Abell 2744	     & ~ & "      &&  8557  & ACIS-I & VFAINT & 23.5 & Jun. 14 2007\\
&&&&&&&&\\
\hline \vspace{-0.2cm}\\
\multicolumn{1}{c}{} &
\multicolumn{1}{c}{} &
\multicolumn{1}{c}{} &
\multicolumn{4}{c}{\hspace{2cm} NON-RADIO HALO CLUSTERS} &
\multicolumn{1}{c}{} &
\multicolumn{1}{c}{} \\
\vspace{-0.1cm}
&&&&&&&&\\

Abell 576            & ~ & 0.0381 &&  3289  & ACIS-S & VFAINT & 24.7 & Oct. 27 2002\\
Abell 1795 (1)       & ~ & 0.0634 &&  3666  & ACIS-S & VFAINT & 13.2 & Jan. 18 2004\\
Abell 1795 (2)       & ~ & "      &&  5287  & ACIS-S & VFAINT & 14.3 & Jun. 10 2002\\
Abell 1795 (3)       & ~ & "      &&  5289  & ACIS-I & VFAINT & 9.6  & Jan. 14 2004\\
Abell 2029 (1)       & ~ & 0.0779 &&  891   & ACIS-S & FAINT  & 18.7 & Apr. 12 2000\\
Abell 2029 (2)       & ~ & "      &&  4977  & ACIS-S & FAINT  & 74.8 & Jan. 8  2004\\
Abell 2029 (3)       & ~ & "      &&  6101  & ACIS-I & VFAINT & 8.2 & Dec. 17 2004\\
Abell 478 (1)        & ~ & 0.088  &&  1669  & ACIS-S & FAINT  & 39.9 & Jan. 27 2001\\
Abell 478 (2)        & ~ & "      &&  6102  & ACIS-I & VFAINT & 9.2  & Sep. 13 2004\\
Abell 2204 (1)       & ~ & 0.152  &&  499   & ACIS-S & FAINT  & 10.0 & Jul. 29 2000\\
Abell 2204 (2)       & ~ & "      &&  6104  & ACIS-I & VFAINT & 9.6  & Sep. 20 2004\\
&&&&&&&&\\
\hline
\end{tabular}
\end{center}
\end{table*}

\subsection{Sample Selection}

Our targets are drawn from the sample of X-ray luminous clusters with
large, extended radio halos discussed by Feretti \& Giovannini
(2007). These clusters have high quality X-ray data available on the
Chandra archive. All of the clusters show clear evidence for recent,
major merger activity and none of the clusters exhibits a cooling core
(Feretti \& Giovannini 2007).  In order to minimize systematic
uncertainties associated with modelling the background in the Chandra
observations, we have imposed a lower redshift limit $z>0.03$; this
excludes the Coma Cluster from our study. We also require at least
$2\times10^4$ net counts, which excludes CL\,0016+1609. Our target
list of radio halo clusters includes Abell 665, 2163, 2219, 2255,
2319, 2744, and the Bullet Cluster 1E\,0657-56.
 
As a control, we have also observed five comparably X-ray luminous
clusters that are known $not$ to contain radio halos: Abell 478, 576,
1795, 2029 and 2204 (Feretti \& Giovannini 2007). Most are highly
relaxed and contain central cooling cores, although Abell 576 exhibits
a more active dynamical state (\eg Dupke \etal 2007; Kempner \& David
2004a).

\subsection{Data Reduction}

The Chandra observations were carried out using the Advanced CCD
Imaging Spectrometer (ACIS) between March 2000 and June 2007.  The
standard level-1 event lists produced by the Chandra pipeline
processing were reprocessed using the $CIAO$ (version 4.1.1) software
package, including the appropriate gain maps and updated calibration 
products.\footnote{Our analysis uses the latest $Chandra$ calibration
products released in January 2009 as incorporated into $CALDB$ version
4.1.1.  The results are robust to residual systematic uncertainties in the 
calibration (see section 3.1).}
Bad pixels were removed and standard grade
selections applied. Where possible, the extra information available in
VFAINT mode was used to improve the rejection of cosmic ray
events. The data were cleaned to remove periods of anomalously high
background using the standard energy ranges and binning methods
recommended by the Chandra X-ray Center.  The net exposure times after
cleaning are summarized in Table~\ref{table:obs}. 
Separate photon-weighted response matrices
and effective area files were constructed for each region analyzed.

\subsection{Spatially-resolved spectroscopy}

\subsubsection{Contour binning}

The individual regions for the spectral analysis were determined using
the contour binning method of Sanders (2006), which groups
neighboring pixels of similar surface brightness until a desired
signal-to-noise threshold is met. For data of the quality discussed
here, regions are frequently small enough that the X-ray emission from
each can be approximated usefully by a single temperature plasma
model.  We consider only those regions of the clusters within which
the spectra are determined reliably, with conservative allowances for
systematic uncertainties in the background subtraction.  In detail, we
set the outermost radius to be the radius at which the ratio of
background-to-total counts in the $0.8-7.0$ keV energy range rises to
20 per cent. At this surface brightness threshold, a 10 per cent shift
in the background normalization will result in an approximately 10 per
cent shift in the temperature of an isothermal plasma model, for both
the ACIS-S and ACIS-I detectors.  The total number of spectral regions
per cluster ranges from three to almost 100.  Regions vary in size from a
few hundred to approximately ten thousand square arcsec.

For our standard analysis described in Section 3, each independent
spectral region contains approximately $10^4$ counts.  This is
sufficient to determine both the metallicity of the cluster gas and
the photon index of a power-law component, where present, to
approximately 10 per cent accuracy.  The temperature and density of
the cluster gas are always determined to higher precision.

In order to generate the detailed thermodynamic maps of temperature,
pressure and entropy discussed in Section 4, we have also carried out
a second analysis using a finer binning of $\sim3,000$
counts/region. For this second analysis, no variable power-law emission
components were included and the metallicity was fixed at the
appropriate emission-weighted mean values determined from the
$\sim10^4$ counts-per-bin results.  The analysis with the finer regions
is sufficient to determine the temperature to $\sim$10 per cent 
accuracy for 10 keV gas, or $\sim3$ per cent accuracy for the coolest
regions in the cores of the dynamically relaxed, non-radio halo
clusters.

\subsubsection{Background modelling}

Searching for non-thermal emission signatures in Chandra spectra
requires careful modelling of the astrophysical and instrumental
background emission. We have taken care to account for spatial and
temporal variations in these components.

Background spectra for the appropriate detector regions were extracted
from the blank-sky fields available from the Chandra X-ray
Center. These were normalized by the ratio of the observed and
blank-sky count rates in the 9.5-12 keV band (the statistical
uncertainty in the observed $9.5-12$ keV flux is less than 5 per cent
in all cases).  

To enable further refinement of the background models, for each ACIS-S
observation, additional spectra were extracted from source-free
regions of Chip 5; for ACIS-I observations, additional spectra were
extracted from regions of Chips 0-3, as far as possible from the
cluster. These spectra were compared with the blank-sky fields in
order to check for consistency and, in particular, to identify the
presence of additional soft X-ray emission components along the line
of sight, and/or scattered solar X-rays.  Where excess soft background
emission was detected, this was modelled and appropriate additional 
components
included in the spectral analysis described below.  Additional soft
background emission components were required in the analysis of the
July 2002 observation of 1E\,0657-56 and for each observation of Abell
2029, 2163, and 2204. Abell 2029 in particular lies behind the North
Polar Spur and has clear excess soft background emission visible in
the Chandra spectra.

\subsubsection{Basic Spectral Analysis}

\begin{table*}
\begin{center}
\caption{Summary of results from the analysis of the $10^4$ count
regions for the merging, radio halo clusters using spectral models
A-D.  $\Delta C$ and $\Delta \nu$ are the improvements in C-statistic
and change in the number of degrees of freedom with respect to the
default model (A, or in the case of Abell 2163, B) for each subsequent
fit.  The value $P$ gives the approximate probability that such an
improvement would occur randomly. (Low $P$ values indicate significant
improvements.) $\Gamma$ is the photon index of the power-law
components in model C.  $kT_2$ is the temperature of the
additional MEKAL component in model D. 
}
\label{table:rhc} 
\begin{tabular}{ c c c c c }
Model & A & B & C & D \\
\hline
A2319 &(10 regions)\\
$\Delta$C & - & 18.5 & 75.6 & 61.8\\
$\Delta\nu$ & - & 1 & 10 & 11 \\
$N_H$ & 8.09 & 6.8$^{+0.3}_{-0.3}$ & 8.09 & 8.09\\
$\Gamma$ & - & - & 2.0 & -\\
$kT_2$ & - & - & - & $80^{+0}_{-8}$ \\
P & - & $1.8\times10^{-5}$ & 4.8$\times10^{-12}$ 
& $4.9\times10^{-9}$ \\
\hline
A2255 &(4 regions)\\ 
$\Delta$C & - & 10.1 & 25.9 & 28.3\\
$\Delta\nu$ & - & 1 & 4 & 5 \\
$N_H$ & 2.50 & 1.1$^{+0.4}_{-0.4}$ & 2.50 & 2.50\\
$\Gamma$ & - & - & 2.0 & -\\
$kT_2$ & - & - & -  & $80^{+0}_{-40}$ \\
P & - & $1.0\times10^{-3}$ & $3.6\times10^{-5}$ 
& $3.5\times10^{-5}$ \\
\hline
A665 &(3 regions)\\ 
$\Delta$C & - & 21.5 & 25.5 & 14.3\\
$\Delta\nu$ & - & 1 & 3 & 4 \\
$N_H$ & 4.33 & 0.6$^{+0.8}_{-0.6}$ & 4.33 & 4.33\\
$\Gamma$ & - & - & 2.0 & -\\
$kT_2$ & - & - & - & $18^{+31}_{-4}$ \\
P & - & 3.9$\times10^{-6}$ & 1.4$\times10^{-5}$ 
& $7.0\times10^{-3}$ \\
\hline
A2163 &(17 regions)\\
$\Delta$C & - & - & 54.1 & 22.2\\
$\Delta\nu$ & - & - & 17 & 18\\
$N_H$ & - & 15.0$^{+0.3}_{-0.4}$ & 18.3$^{+0.8}_{-1.0}$
& $15.5^{+0.3}_{-0.4}$\\
$\Gamma$ & - & - & 2.0 & - \\
$kT_2$ & - & - & - & $60^{+20}_{-20}$ \\
P & - & - & $1.0\times10^{-5}$ & $0.224$\\
\hline
\end{tabular}
\end{center}
\end{table*}

The spectra have been analyzed using XSPEC (version 11.3; Arnaud
1996), the MEKAL plasma emission code (Kaastra \& Mewe 1993), and the
photoelectric absorption models of Balucinska-Church \& McCammon
(1992). All spectral fits were carried out in the $0.6-7.0$
keV energy band. The extended C-statistic available in XSPEC, 
which allows for background subtraction, was used for all fitting.

The default spectral model (model A) applied to each spatial region
consists of a MEKAL plasma model at the redshift of the cluster with
the absorbing column density along the line of sight fixed at the
Galactic value determined from HI studies (Kalberla \etal
2005).\footnote{A comparison of the line-of-sight Galactic column
densities to the target clusters determined by the HI study of
Kalberla \etal (2005) with the earlier work of Dickey \& Lockman
(1990) shows good overall agreement, with one notable exception: the
Galactic column density for 1E\,0657-56 is $6.53 \times 10^{20}$ atom
cm$^{-2}$ in the Dickey \& Lockman (1990) study, whereas it is
$4.89\times 10^{20}$ atom cm$^{-2}$ in the later Kalberla \etal (2005)
work.  We have adopted the Kalberla \etal (2005) result as the more
accurate value.}  The normalization and temperature in each spatial
region are included as free parameters in the fits. For the initial
analysis with $10^4$ counts per bin, the metallicity is also allowed
to vary independently in each region.

A series of more sophisticated spectral models were also examined and
the results compared to those obtained with model A: model B is the
same as model A but with the line-of-sight absorption allowed to vary
as a single, additional free parameter (linked to a common value
across all regions). Model C has Galactic absorption but introduces
an additional power-law
component into each spectral region: this approximates the
spectrum of a range of possible non-thermal or quasi-thermal 
emission components.  The
normalization of the power-law component in each region is a separate
free parameter. 
The photon index of the
power-law component is initially fixed 
to a canonical value of $\Gamma=2.0$. (We later
examined models in which the photon index of the power-law
component was allowed to vary, either in unison across all regions or
in each spatial region independently, see section 3.2).

Finally, model D
builds on model A, but introduces a second MEKAL component in each
region, with the temperature of the second emission component linked
to vary in unison across all regions. For computational reasons, an
upper limit on the temperature of this component of 80 keV was
imposed.  Model D also has the column density fixed at the Galactic value.

Due to degeneracies between parameters for models C and D,
we constrain the metallicity in each region to remain within $\pm50$ per cent
of the values determined with model A.  By doing this and keeping 
the absorbing column density fixed at the Galactic value, we limit the
impact of unphysical model solutions.

For Abell 478 and Abell 2163, the line-of-sight absorbing column
density is known to differ substantially from the radio HI values of $1.5
\times 10^{21}$\apc and $1.09 \times 10^{21}$\apc, respectively
(Johnstone \etal 1992; Allen \etal 1993; Pointecouteau \etal 2004;
Elbaz \etal 1995).  For these clusters, model B is therefore adopted
as the default model and models C and D also have the column density
included as a free parameter. We caution
that for Abell 478, the column density is also known to vary as a
function of position (Allen \etal 1993). For this reason and for this
cluster alone, the column density was allowed to vary independently in
each region studied.

\section{Results}

\subsection{The statistical evidence for non-thermal-like 
components}
\label{section:evidence}

\addtocounter{table}{-1}
\begin{table*}
\begin{center}
\caption{Continued.}
\begin{tabular}{ c c c c c }
Model & A & B & C & D \\
\hline
A2219 &(8 regions)\\ 
$\Delta$C & - & 0.02 & 3.3 & 3.9\\
$\Delta\nu$ & - & 1 & 8 & 9 \\
$N_H$ & 1.76 & 1.8$^{+0.3}_{-0.3}$ & 1.76 & 1.76\\
$\Gamma$ & - & - & 2.0 & -\\
$kT_2$ & -& - & - & $0.4^{+0.4}_{-0.2}$ \\
P & - & 0.89 & 0.91 & 0.92\\
\hline
1E\,0657-56 &(55 regions)\\ 
$\Delta$C & - & 10.3 & 107.6 & 88.8\\
$\Delta\nu$ & - & 1 & 55 & 56 \\
$N_H$ & 4.89 & 4.44$^{+0.15}_{-0.18}$ & 4.89 & 4.89\\
$\Gamma$ & - & - & 2.0 & -\\
$kT_2$ & - & - & - & $80^{+0}_{-20}$ \\
P & - & $1.3\times10^{-3}$ & $2.9\times10^{-5}$ 
& $3.4\times10^{-3}$ \\
\hline
A2744 &(6 regions)\\ 
$\Delta$C & - & 0.3 & 7.1 & 17.0\\
$\Delta\nu$ & - & 1 & 6 & 7 \\
$N_H$ & 1.39 & 1.4$^{+0.4}_{-0.4}$ & 1.39 & 1.39\\
$\Gamma$ & - & - & 2.0 & -\\
$kT_2$ & - & - & - & $22^{+18}_{-9}$ \\
P & - & 0.58 & 0.31 & 0.017\\
\hline
\end{tabular}
\end{center}
\end{table*}

\begin{table*}
\begin{center}
\caption{
Details as for table \ref{table:rhc} but for the relaxed, non-radio
halo clusters.  For Abell 478, the line of sight Galactic column
density is known to significantly exceed the value determined from HI
studies (Johnstone \etal 1992; Allen \etal 1993) and vary spatially
across the cluster.  For this cluster, 
the $N_{\rm H}$ value was allowed to vary independently
in each region.  Low $P$ values indicate significant improvements.
For model D, only regions beyond a
radius of $100h_{70}^{-1}$ kpc were used, to limit the impact of
projection effects in the presence of strong temperature gradients
(Section 3.2).
}
\label{table:nrhc} 
\begin{tabular}{ c c c c c }
Model & A & B & C & D \\
\hline
A576 &(4 regions)&&&(4 regions)\\ 
$\Delta$C & - & 0.6 & 7.5 & 4.5\\
$\Delta\nu$ & - & 1 & 4 & 5 \\
$N_H$ & 5.50 & 5.5$^{+0.3}_{-0.4}$ & 5.50 & 5.50\\
$\Gamma$ & - & - & 2.0 & -\\
$kT_2$ & - & - & - & $15^{+65}_{-11}$ \\
P & - & 0.44 & 0.11 & 0.48\\
\hline
A1795 &(24 regions)&&&(8 regions)\\ 
$\Delta$C & - & 0.08 & 31.4 & 24.5\\
$\Delta\nu$ & - & 1 & 24 & 9 \\
$N_H$ & 1.19 & $1.18^{+0.09}_{-0.18}$ & 1.19 & 1.19 \\
$\Gamma$ & - & - & 2.0 & - \\
$kT_2$ & - & - & - & $80^{+0}_{-60}$ \\
P & - & 0.77 & 0.14 & $3.7\times10{-7}$ \\
\hline
A2029 &(90 regions)&&&(40 regions)\\ 
$\Delta$C & - & 9.1 & 113.0 & 68.8\\
$\Delta\nu$ & - & 1 & 90 & 41\\
$N_H$ & 3.26 & 3.03$^{+0.06}_{-0.07}$ & 3.26 & 3.26\\
$\Gamma$ & - & - & 2.0 & -\\
$kT_2$ & - & - & - & $80^{+0}_{-16}$ \\
P & - & $2.6\times10^{-3}$ & 0.051 & $4.2\times10{-3}$\\
\hline
A478 &(37 regions)&&&(21 regions)\\
$\Delta$C & - & - & 45.7 & 28.3\\
$\Delta\nu$ & - & - & 37 & 22\\
$N_H$ & - & $24.7-31.1$ & $24.7-35.8$ & $24.6-31.3$\\
$\Gamma$ & - & - & 2.0 & -  \\
$kT_2$ & - & - & - & $41^{+39}_{-18}$\\
P & - & - & 0.15 & 0.17\\
\hline
A2204 &(7 regions) &&&(3 regions)\\ 
$\Delta$C & - & 6.6 & 1.5 & 5.2\\
$\Delta\nu$ & - & 1 & 7 & 4\\
$N_H$ & 5.67 & 6.4$^{+0.3}_{-0.3}$ & 5.67 & 5.67\\
$\Gamma$ & - & - & 2.0 & -\\
$kT_2$ & - & - & - & $3.1^{+1.5}_{-1.2}$ \\
P & - & 0.01 & 0.98 & 0.27\\
\hline
\end{tabular}
\end{center}
\end{table*}

\begin{figure} 
\vspace{0.2cm}
\hbox{
\hspace{-0.2cm}
\includegraphics[width=0.45 \textwidth]{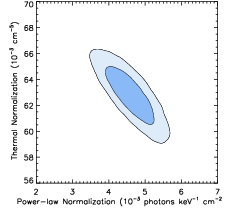}
}
\caption{The 68 and 95 per cent confidence contours of the power-law and
thermal normalizations summed across all regions in Abell 2319.  Contours
were determined from a Markov Chain Monte Carlo analysis and confirm that
the power-law component is detected (normalization $>$ 0) at high significance.
}
\label{fig:plnorm}
\end{figure}

Tables \ref{table:rhc} and \ref{table:nrhc} summarize the results from
the fits with spectral models A-D for the radio halo clusters and the
control sample of non-radio halo systems, respectively. The tables
list the total decrease in C-statistic ($\Delta C$) summed across all
regions in each cluster, and the increase in the number of degrees of
freedom ($\Delta \nu$) for each model with respect to the default
model (typically model A, but model B for Abell 478 and 2163),
together with the approximate probability, $P$, that such an
improvement could be obtained randomly.\footnote{To estimate the
probability that the improvements to the fits could be obtained by
random chance, we have calculated a pseudo $F-$statistic, formed by
the ratio $F=\Delta C/\Delta \nu$. This is similar to the usual
$F-$statistic formed with the chi-square estimator
(\eg Bevington 1969), and is expected to
provide useful estimates of the probability for cases where the
spectral model provides a reasonable description of the data.} The
tables also list the best-fitting absorbing column density for cases
where this is included as a free parameter.

Comparing the results for the simplest single-temperature models, A
and B, we see that for four out of six radio halo clusters for which
both models were examined, allowing the absorbing column to fit freely
leads to a significant improvement in the fit ($\Delta C \geq
10$ for 1 additional degree of freedom). The most significant
improvement, $\Delta C \sim 22$, is obtained for Abell 665.
In all four of these six cases, however, the best-fit
column density is significantly $lower$ than the Galactic value
determined from HI studies, a result that is physically implausible
and which suggests that model B does not provide a complete
description of the spectra. (For column densities $\approxlt
8\times10^{20}$ atom cm$^{-2}$, the Galactic HI values are expected to
be accurate to better than 10 per cent). 
The only two radio halo clusters for which
significant improvements ($\geq$99 per cent confidence) to the fits
were not obtained using model B over model A are Abell 2219 and Abell
2744. 
For the seventh radio halo
cluster, Abell 2163, which lies at low Galactic latitude, an accurate
Galactic column density estimate is not available from HI studies and
model A was therefore not examined; model B is taken as the default
model for this cluster. 

Importantly, $none$ of the control sample of non-radio halo clusters
show a comparably significant improvement when using model B over
model A.\footnote{We have confirmed that consistent results are
obtained whether the ACIS-S and ACIS-I data for individual clusters
are analyzed separately or together.}  As discussed above, Abell 478
is excluded from this test since it lies at low Galactic latitude and
the estimate of the line-of-sight column density from HI studies is
poorly determined.  The most statistically significant detection is
for Abell 2029, although the background model for this cluster is
complex with relatively high systematic uncertainties (Section
2.3.2). It is unlikely that our modelling of the excess soft emission
along the line of sight, associated with the North Polar Spur, is
correct in detail (we assume it to have constant surface brightness
across the field).  Note also that the data for this cluster exhibit
an exceptionally high signal to noise ratio and the best fit column
density is only lower than the Galactic value by $\sim10^{19}$ atom
cm$^{-2}$.

Comparing the results for models C and A, we see that the introduction
of a power-law component with photon index $\Gamma=2.0$ leads to
comparable or 
even larger improvements in the fits to the same four radio
halo clusters (\ie those that fitted to sub-Galactic column densities
with model B), as well as for Abell 2163, for which a reliable Galactic 
column density
measurement is not available. 
For Abell 2219, 2744, and for $all$ of the relaxed non-radio halo
clusters, however, no comparably 
significant improvement in the fits is obtained with model C over model A.
Figure \ref{fig:plnorm} shows the combined power-law normalization versus
thermal normalization (in units of $10^{-3}$) summed
across all regions in Abell 2319.  Power-law emission is detected at high 
significance.  The power-law normalization is approximately 10 per cent
of the thermal normalization and, thus also, approximately 10 per cent
of the total flux.

We conclude, based on the improvements obtained with models B and C,
that 5/7 radio halo clusters show evidence for additional model
components, over and above those accounted for by the default thermal model.
\footnote{Similar conclusions regarding the presence or otherwise of
power-law components in the radio halo clusters are obtained when the
photon index of the power-law component is also included as a single
free parameter.}  The additional emission components can be
approximated by simple power-law models, although we stress that these
models are unlikely to provide a complete, physical description of the
data.  In contrast, the relaxed, non-radio halo systems do not show
comparable evidence for such power-law components.

\begin{figure*}
\vspace{0.2cm}
\hbox{
\hspace{0.2cm}
\includegraphics[width=0.45 \textwidth]{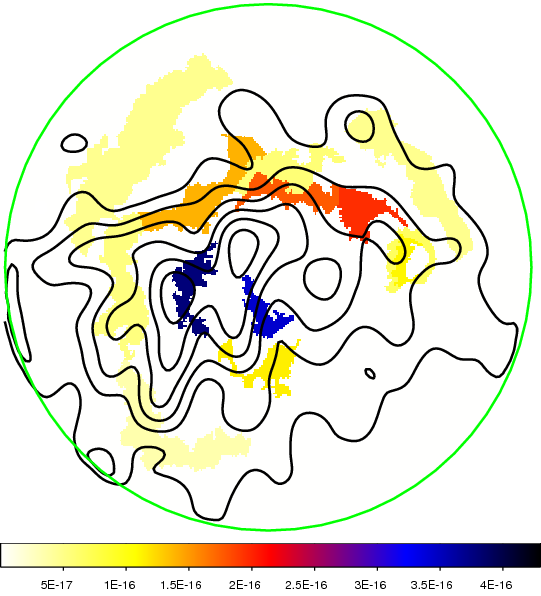}
\hspace{1.0cm}
\includegraphics[width=0.45 \textwidth]{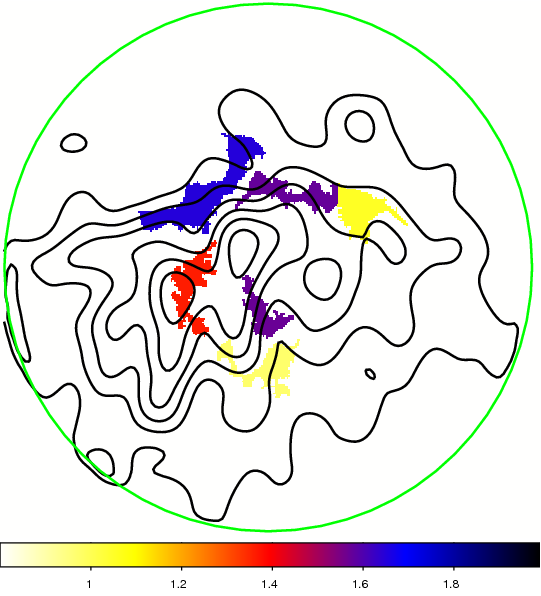}
}
\caption{(a) left panel: Spatial map of the surface brightness (in
erg\,cm$^{-2}$s$^{-1}$arcsec$^2$) of non-thermal-like X-ray emission
in the Bullet Cluster, 1E\,0657-56 ($z=0.297$).  A power-law
component is only included in regions where it is statistically
required at greater than 90 per cent significance (Section
\ref{section:spatial}).  The 1.34 GHz radio surface brightness
contours from Liang \etal (2000) are overlaid in black. (Radio point
sources have been removed).  (b) right panel: Map of the photon index of
the power-law components, with the radio surface brightness contours
overlaid.  Only regions with `non-thermal' 
surface brightness greater than $10^{-16}$ 
ergs s$^{-1}$ cm$^{-2}$ arcsec$^{-2}$ are shown in this map.
Each spatial region has $\sim10^4$ net counts in the
$0.6-7.0$ keV Chandra band. }
\label{fig:1percent}
\end{figure*}

Similar conclusions on the presence or otherwise of non-thermal-like
emission in the radio halo clusters are also drawn from the fits
with model D, where a second MEKAL component, rather than a
power-law model, is incorporated into the modelling (other than for
Abell 2163; although this cluster
also has the column density included as a free parameter).
The nominal temperatures of the additional
emission components are typically high, with $kT_2\approxgt15$ keV.

As discussed in Section 3.2, the majority of the control sample of
relaxed, non-radio halo clusters (all, excluding Abell 576) are known
to contain strong temperature gradients in their (cooling) cores.  The
presence of such gradients will naturally lead to requirements
for additional temperature components in fits to projected spectra for
the central regions of these systems. To mitigate this effect, for the
analysis of the control sample of clusters with model D, we have
excluded the inner $100h_{70}^{-1}$kpc radius. (The reduced number of
regions studied are indicated in column 4 of Table 3.) Of the control
sample, Abell 2029 presents a mild requirement for additional
temperature components, and Abell 1795 a more significant requirement,
with most of the signal arising from within radii $\sim 200h_{70}^{-1}$kpc
in both cases. Possible origins for these requirements 
are discussed in Section 5.

Finally, to gauge the level at which residual instrument calibration
issues may affect the results, we have repeated the analysis in three
other energy bands: $0.6-4.0$ keV, $0.6-5.0$ keV and $2.0-7.0$ keV.
For the merging, radio halo clusters, similar detections of additional
emission components are made using each energy band.
The control sample of regular, non-radio halo
clusters provides comparable null results in all energy bands studied.

\begin{table*}
\begin{center}
\caption{Results from the spatial mapping of the power-law components
in the radio halo clusters (Section \ref{section:spatial}).  Column 2
lists the Galactic column density for each cluster determined from HI
studies (Kalberla \etal 2005; for Abell 2163 a reliable column density
measurement is not available from radio data). 
Columns 3 lists the total number of regions for
which a power-law component is detected at 90 per cent significance or
greater. Columns $4-7$ list the column density (in units of
$10^{20}$\apc) determined from the fits, the mean photon index of the
required power-law components, and the summed $0.6-7.0$ keV flux (in
units of $10^{-12}$ ergs s$^{-1}$ cm$^{-2}$) of these components,
respectively. (Regions with power-law photon indices, $\Gamma<1$, 
typically exhibit very low non-thermal surface brightness. We consider 
such results to be unphysical and have excluded them from the average
values in columns 5,6).
}
\label{table:exclude}
\begin{tabular}{ c c c c c c c c c c c c c } 
\multicolumn{1}{c}{} &
\multicolumn{1}{c}{} &
\multicolumn{1}{c}{} &
\multicolumn{1}{c}{} & \\
\multicolumn{1}{c}{Cluster} &
\multicolumn{1}{c}{} &
\multicolumn{1}{c}{$N_{\rm H,Gal}$} &
\multicolumn{1}{c}{} &
\multicolumn{1}{c}{\# regions} &
\multicolumn{1}{c}{$N_{\rm H}$} &
\multicolumn{1}{c}{$\Gamma_{\rm mean}$} &
\multicolumn{1}{c}{PL Flux} \\
Abell 2319 &~& 8.09 &~& 3/10 & $8.3^{+0.4}_{-0.4}$ & $2.0^{+0.4}_{-0.3}$ & $11^{+8}_{-5}$ \\
Abell 2255 &~& 2.50 &~& 2/4 & $3.01^{+1.0}_{-0.7}$ & $1.68^{+0.19}_{-0.27}$ & $5.1^{+1.2}_{-1.2}$ \\
Abell 665 &~& 4.33 &~& 2/3 & $2.6^{+1.2}_{-1.5}$ & $1.63^{+0.10}_{-0.21}$ & $4.2^{+1.4}_{-1.2}$ \\
Abell 2163 &~& - &~& 4/17 & $15.4^{+0.4}_{-0.3}$ & $1.51^{+0.05}_{-0.05}$ & $3.9^{+1.0}_{-1.0}$ \\
Abell 2219 &~& 1.76 &~& 0/8 & $1.7^{+0.3}_{-0.3}$ & - & - \\
Abell 2744 &~& 1.39 &~& 3/6 & $1.7^{+0.4}_{-0.4}$ & $1.66^{+0.40}_{-0.13}$ & $0.4^{+0.2}_{-0.3}$ \\
1E\,0657-56 &~& 4.89 &~& 13/55 & $4.99^{+0.15}_{-0.14}$ & $1.50^{+0.05}_{-0.05}$ & $0.95^{+0.10}_{-0.11}$ \\

\end{tabular}
\end{center}
\end{table*}

\subsection{Detailed spatial mapping of the `non-thermal' components}
\label{section:spatial}

The results of Section 3.1 provide a simple, statistical measure of
the overall significance of non-thermal-like emission
signatures in the clusters. However, even where these overall
signatures are strong, as in 5/7 of the 
radio halo clusters, we do not necessarily expect power-law
emission components to be present in every region in the clusters. We
have therefore attempted to map the spatial variation of this
emission.

Starting with model A, a single temperature model with the absorption
fixed at the Galactic HI value, we have determined, for each
individual region of each cluster, the statistical significance of the
improvement to the fit obtained by introducing a power-law component
with a photon index that is allowed to fit freely in that region.
Where the significance of the improvement was found to exceed a
threshold value (we adopt 90 per cent) the power-law component was
kept in the model for that region.\footnote{We have constrained
the metallicity to lie within $\pm50$ per cent of the value determined
with model A.  Similar results are also obtained if the metallicity is
simply constrained to be less than solar.}  Where it was not, model A
(single, thermal emission component only) was used as the appropriate
region model. In this way, the number of regions requiring power-law 
emission components in each cluster were determined. Finally, the
overall absorbing column density was allowed to vary and checked
for consistency against Galactic HI survey results.\footnote{ An
alternative algorithm to identify regions requiring power-law
components was also examined, wherein the column density was allowed
to fit freely in each region, although constrained not to vary by more
than 50 per cent from the Galactic value. This algorithm led to 
similar results on the regions requiring power-law emission
components.}  Maps of the surface brightness of the power-law
emission, where detected, and the photon index can then be generated
in a straightforward manner. By including `non-thermal' emission
components only where they are statistically required, we minimize the
impact of random errors and residual calibration uncertainties on the
maps.

The best fit absorption column densities and mean emission-weighted
photon indices determined from the detailed mapping at the 90 per cent
threshold are summarized in Table \ref{table:exclude}. In all cases,
the measured column densities are consistent with the Galactic values
(Kalberla \etal 2005) at the $\sim 95$ per cent (2$\sigma$) confidence
level. This contrasts with the results obtained with model B (single
temperature model) where column densities less than the Galactic value
was observed in most cases (Section 3.1).\footnote{We have also mapped
the spatial variation of the non-thermal-like emission using a more
restrictive threshold set at 99 per cent confidence.  The best-fit
column densities and emission-weighted photon indices obtained from
this analysis are broadly consistent with those determined using the
90 per cent threshold, although the column density for 1E\,0657-56 is
reduced slightly below the Galactic value.} 

Excluding regions with unphysically low photon indices \ie $\Gamma<1$,
which are almost always detected at very low surface brightness and
which can be plausibly associated with residual calibration and
background subtraction errors, the mean emission-weighted photon
indices for the clusters range from $1.5 < \Gamma < 2.0$.

The results for the Bullet Cluster, 1E0657-56, which has the longest
Chandra exposure 
components are shown in Fig \ref{fig:1percent}a.  Also shown overlaid
in Fig \ref{fig:1percent}a are the contours of constant radio halo
brightness from the 1.34 GHz radio data of Liang \etal (2000; radio
point sources have been removed). The correspondence between the radio
contours and power-law X-ray emission map is interesting. The peaks of
the radio halo emission and power-law X-ray emission both occur in the
main cluster center.  There is also an extension of both emission
components, albeit at lower surface brightness, toward the merging
`bullet' subcluster. We conclude that there is a correspondence
between the locations of the strongest power-law X-ray emission and
strongest non-thermal radio emission in the cluster.

\begin{figure}
\hbox{
\hspace{0.0cm}
\includegraphics[width=0.47 \textwidth]{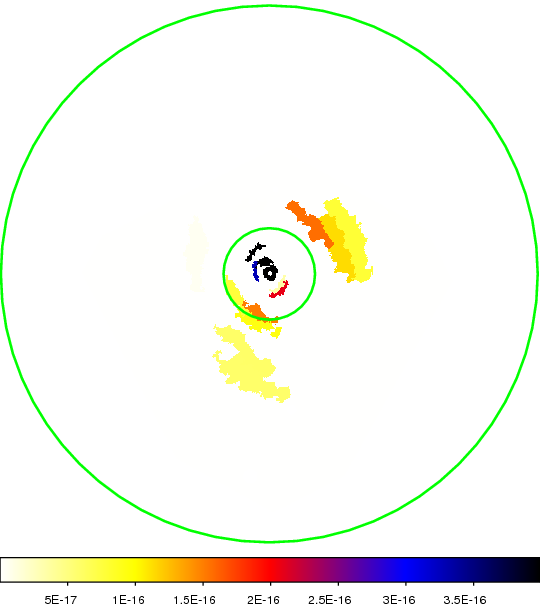}
}
\caption{Spatial map of the surface brightness (in units of erg s$^{-1}$ 
cm$^{-2}$ arcsec$^{-2}$ of non-thermal-like X-ray emission in Abell 2029.
For comparison with Fig. 1, the scale has been
adjusted to represent what would been seen if this cluster were at the
same redshift as 1E\,0657-56 and the observation time were extended to
maintain the current signal-to-noise ratio.  The inner circle marks a
radius of $100h_{70}^{-1}$ kpc, within which the observed temperature
declines steeply and projection effects are strong.  The outer circle
shows marks the outer limit of the analysis. }
\label{fig:sba2029}
\end{figure}

Fig \ref{fig:1percent}b shows the spatial variation of the photon
index of the power-law X-ray emission in 1E\,0657-56, again with the
radio contours overlaid. The regions of strongest power-law flux have
a photon index $\Gamma \sim 1.4-1.8$. Regions with photon index
$\Gamma<1.0$, all of which are detected at relatively low surface
brightness, $SB_{PL}<1\times10^{-16}$, are probably
unphysical. Summing the total flux of all power-law components
detected at a confidence level of 90 per cent or greater and with a
photon index $\Gamma \geq 1$, we obtain a total $0.6-7.0$ keV power
law flux from 1E\,0657-56 of $0.95^{+0.10}_{-0.11}\times10^{-12}$ ergs
s$^{-1}$ cm$^{-2}$ and an emission-weighted photon index of
$1.50^{+0.05}_{-0.05}$.  Extrapolating this result to the $20-100$ keV
band, we predict a flux of $3.8^{+1.9}_{-0.9}\times10^{-12}$ ergs
s$^{-1}$ cm$^{-2}$, consistent with the result of Petrosian
\etal (2006) of $5\pm3\times 10^{-12}$ ergs s$^{-1}$ cm$^{-2}$ based
on RXTE and XMM-Newton X-ray data.

\begin{figure*}
\hbox{
\hspace{0.0cm}
\includegraphics[width=0.47 \textwidth]{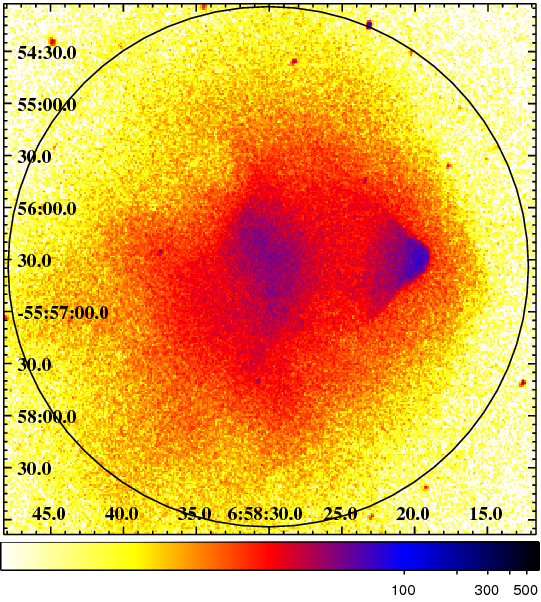}
\hspace{0.9cm}
\includegraphics[width=0.47 \textwidth]{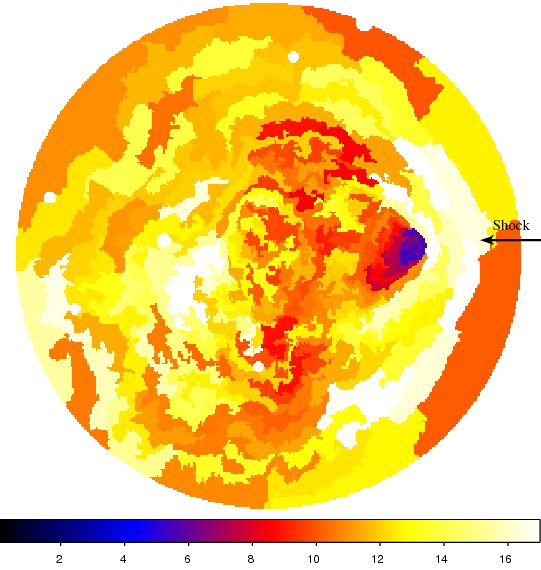}
}
\hbox{
\hspace{0.0cm}
\includegraphics[width=0.47 \textwidth]{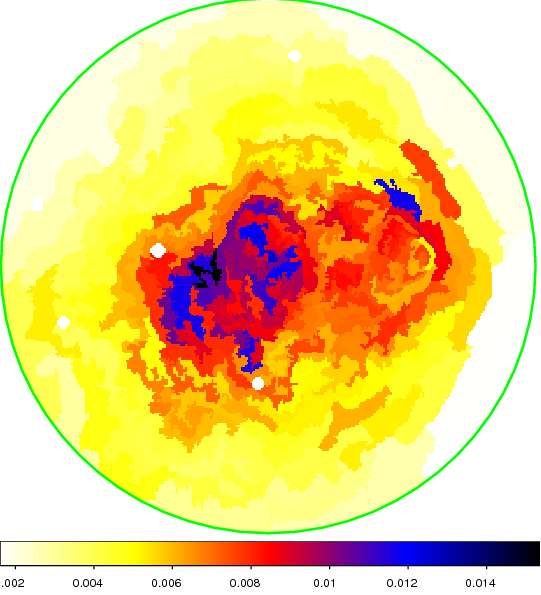}
\hspace{0.9cm}
\includegraphics[width=0.47 \textwidth]{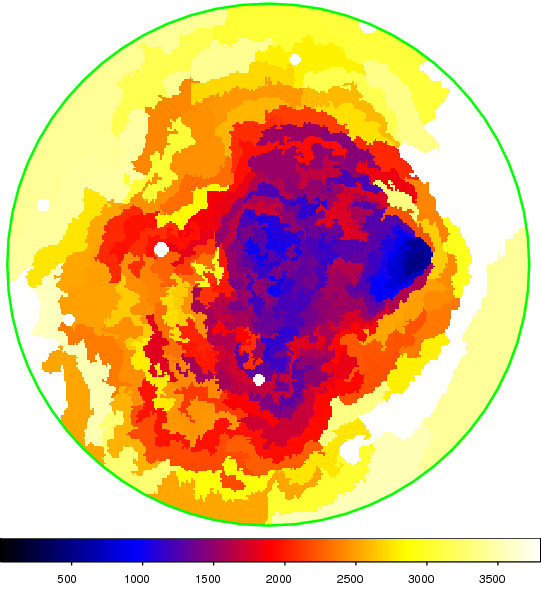}
}
\caption{Thermodynamic maps for the Bullet Cluster, 1E\,0657-56. 
Regions follow lines of constant surface brightness and have $\sim3000$ net
counts, leading to 1$\sigma$ fractional uncertainties in the plotted
quantities of $\approxlt 10$ per cent.  Upper left panel: surface 
brightness (counts) in the 0.8 to 7.0 keV band.  Upper right panel:
temperature, kT, in units of keV determined with spectral model A.
Lower left panel: pressure, P, in units of keV
cm$^{-\frac{5}{2}}$ arcsec$^{-1}$.  Lower right panel:
entropy, S, in units of keV cm$^{\frac{5}{3}}$ arcsec$^{\frac{2}{3}}$.
The shock front analyzed by Markevitch \etal 2002 (see also Markevitch 2006
and Markevitch and Vikhlinin 2007) is marked by the arrow on
the temperature map.
}
\label{fig:bulletT}
\end{figure*}

\begin{figure*}
\hbox{
\hspace{0.0cm}
\includegraphics[width=0.47 \textwidth]{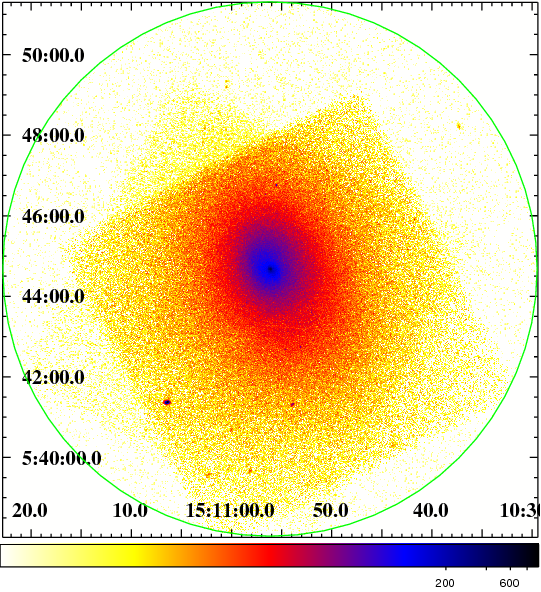}
\hspace{0.9cm}
\includegraphics[width=0.47 \textwidth]{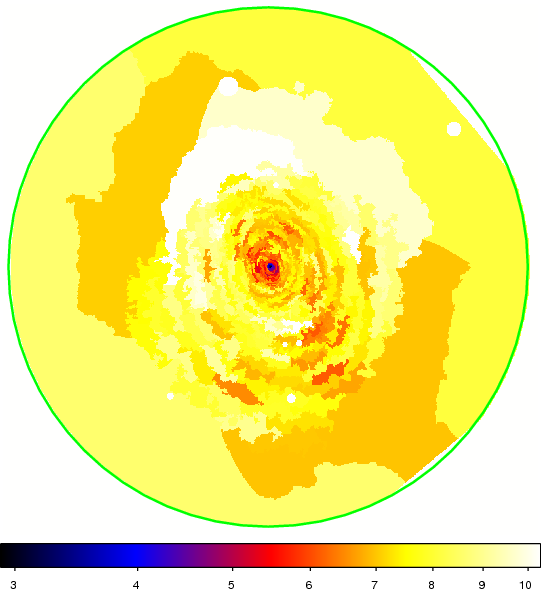}
}
\hbox{
\hspace{0.0cm}
\includegraphics[width=0.47 \textwidth]{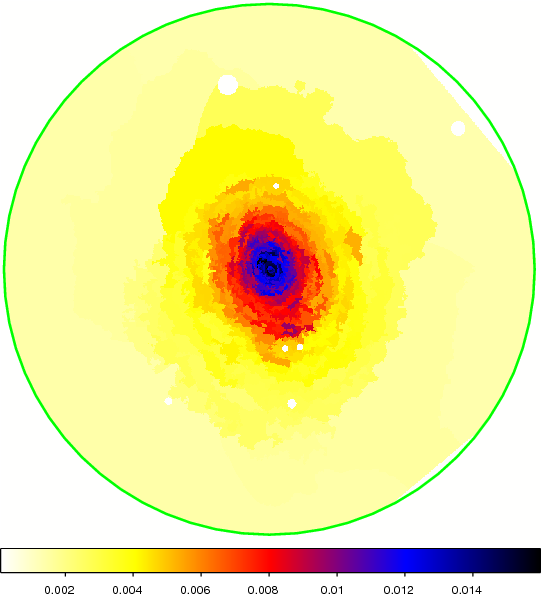}
\hspace{0.9cm}
\includegraphics[width=0.47 \textwidth]{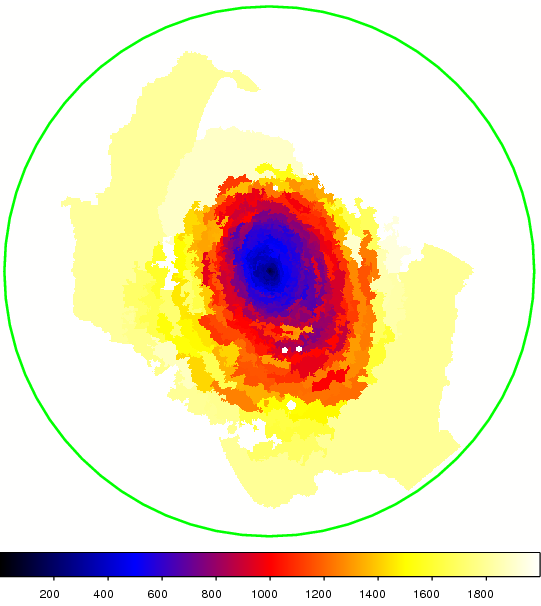}
}
\caption{Thermodynamic maps for the relaxed, non-radio halo cluster Abell 2029. 
Other details as for Fig \ref{fig:bulletT}.
}
\label{fig:a2029}
\end{figure*}

We have applied a similar mapping analysis to our control sample of
non-radio halo clusters.  Fig \ref{fig:sba2029} shows the results for
Abell 2029, which has a deep exposure and the most accumulated X-ray
counts of any cluster in the study.  We have scaled the surface
brightness of the power-law emission in Fig. \ref{fig:sba2029} to
the levels that would be detected if the cluster were observed at the
same redshift as 1E\,0657-56 (assuming an ultra-deep observation with
sufficient exposure time to maintain the current signal-to-noise
ratio). For relaxed clusters like Abell 2029 at relatively low
redshifts, it is important to recall that the temperature profiles are
commonly observed to decline steeply in their cores (e.g. Allen,
Schmidt \& Fabian 2001; Vikhlinin \etal 2005).  The analysis of
projected spectra in the innermost regions of such clusters will
therefore always, given good enough data, provide signatures of
multiphase gas, simply due to projection effects.The inner, green circle in
Fig. \ref{fig:sba2029} marks a radius of $100h_{70}^{-1}$kpc, within
which the temperature profile in Abell 2029 is observed to decline
steeply.  The regions with the most significant detections of
power-law components lie within or near this radius; the photon
indices of these components are typically flat, with $\Gamma
\sim1.0-1.5$. (See also Section 5.2).

The results for the other, relaxed non-radio halo clusters show no convincing
evidence for power-law X-ray emission components beyond their inner
regions.

\section{Thermodynamic mapping of 1E\,0657-56 and Abell 2029}

1E\,0657-56 and Abell 2029 have the deepest Chandra exposures of the
radio halo clusters and control sample of non-radio halo clusters,
respectively, and are therefore the systems best-suited to detailed
thermodynamic mapping.  The Chandra X-ray images and maps of
temperature, pressure and entropy for the two clusters are shown in
Figs.  \ref{fig:bulletT} and \ref{fig:a2029}.

\begin{figure}
\hbox{
\hspace{0.0cm}
\includegraphics[width=0.47 \textwidth]{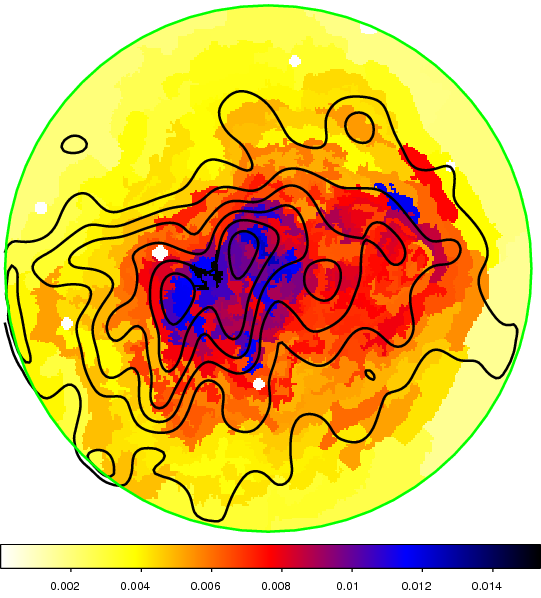}
}
\caption{Pressure map (in keV cm$^{-\frac{5}{2}}$ arcsec$^{-1}$ 
for 1E\,0657-56 with the 
1.34 GHz radio halo contours overlaid. The strongest radio halo 
emission coincides 
with the regions of highest apparent thermal pressure and strongest 
`non-thermal' X-ray emission.}
\label{fig:bulletP}
\end{figure}

The thermodynamic maps for 1E\,0657-56 are complex and confirm that
the cluster is far from hydrostatic equilibrium, having experienced a
recent, massive merger event. The peak of the thermal X-ray surface
brightness is centered on the fast-moving, merging subcluster \ie the
`bullet' (see Markevitch \etal 2002; Markevitch 2006; Markevitch \&
Vikhlinin 2007 and references therein), rather than the overall X-ray
center. The merging subcluster also contains the lowest temperature
and lowest entropy gas. However, the regions of highest thermal
pressure in the cluster are $not$ coincident with the X-ray surface
brightness peak, but are rather found near the overall X-ray
center. The multiple apparent pressure peaks may be sites of recent
and/or ongoing shock activity and might be expected to be locations of
ongoing particle acceleration. Fig. \ref{fig:bulletP} shows the 1.34
GHz radio contours overlaid on the same pressure map. The figure
confirms that the sites of highest thermal pressure and strongest
radio emission are closely correlated (see also Govoni \etal 2004).

The shock front ahead of the merging subcluster in 1E\,0657-56 has
been studied by Markevitch \etal (2002), Markevitch (2006), and
Markevitch \& Vikhlinin (2007).  This shock is also clearly visible in
the thermodynamic maps of Fig. \ref{fig:bulletT}.  Our analysis of the
temperature jump at the shock front
indicates a Mach number $M = 2.5^{+0.5}_{-0.4}$ which corresponds
to a velocity of $v = 4000^{+800}_{-600}$ km s$^{-1}$, consistent with
previous work.
We note a hint of a
second, possible front leading the cool
subcluster in Fig. \ref{fig:bulletT}b, and trailing the main shock
(see also Markevitch \& Vikhlinin 2007).

In contrast, for Abell 2029, the thermodynamic maps, and especially
the pressure map, appear remarkably symmetric and regular.  The
cluster is not in $perfect$ equilibrium, as evidenced by detailed
structure observed in the X-ray image, especially at small radii (see
\eg Clarke, Blanton \& Sarazin 2004), as well as the temperature and
entropy maps shown here (an extension of low entropy gas is observed
$3-4$ arcmin to the southwest). However, the deviations from
elliptical symmetry in the pressure map are small and argue that the
hydrostatic assumption employed in X-ray mass analyses is likely to be
a reasonable approximation in the observed region of the cluster,
confirming earlier conclusions drawn on the basis of the X-ray surface
brightness distribution alone (\eg Buote \& Tsai 1996; Allen \etal
2008).  The peak of the X-ray surface brightness coincides precisely
with the location of the dominant cluster galaxy, IC\,1101. The X-ray
peak also marks the location of the lowest temperature and lowest
entropy gas in the cluster, and of the highest thermal pressure, a
configuration that is convectively stable.

Qualitatively similar results on the thermodynamic properties are
obtained if power-law components consistent with those discussed in
Section 3.2 are accounted for explicitly in the analysis (although the
measured temperatures are, typically, slightly lower). Results
from thermodynamic mapping of other clusters in the sample are
presented in the Appendix.

\section{The origin of the power-law components}

\subsection{Summary of Results}

Our analysis has revealed the presence of significant non-thermal-like
X-ray emission in five of a sample of seven massive, merging clusters
with luminous radio halos.  Our results shed new light on the origin
of this emission. The most important, new facts are 1) the presence of
such components in a large fraction of the radio halo clusters
studied; 2) the absence (or lower detection rate) of such components,
at similar levels, in the control sample of non-radio halo clusters;
3) the observed spectra of the non-thermal-like components, which are
consistent with simple power-law models with relatively flat photon
indices, $1.5<\gamma<2.0$, or bremsstrahlung emission from hot,
$kT>15$ keV, gas; and 4) the spatial correlation between the regions
of the brightest non-thermal radio halo emission, brightest, thermal
X-ray emission, and strongest non-thermal-like X-ray signatures in the
central regions of 1E\,0657-56 (Fig. 1).

Summarizing the statistical evidence from Section 3, we see that the
comparison of results from models B and A provides evidence
($P<10^{-2}$) for excess soft emission (i.e. less-than-Galactic
absorbing column densities) in 4/6 radio halo clusters for which this
test is possible. Of the control sample, only Abell 2029 shows mildly
significant evidence for a soft excess, which can be plausibly
associated with residual uncertainties in the background modelling for
this system.  Comparing results for models C and A, we see that 5/7
radio halo clusters show evidence for power law components (including
the same 4 noted above) whereas, strikingly, 0/5 of the control sample
do. The more detailed mapping discussed in Section 3.2 shows that once
power law components have been included, where statistically required,
the absorbing column densities for the radio halo systems become
consistent with the Galactic values determined from HI studies. The
results from the tests with model D are
somewhat more ambiguous, as might be expected given the additional freedom of 
this spectral
model: 4/7 radio halos show evidence for additional temperature
components, as do 2/5 of the control sample.

\subsection{Physical Explanations}

It has been common in the literature to interpret 
limits on non-thermal X-ray emission from galaxy clusters in terms of
simple IC models, wherein the same population of relativistic electrons
responsible for the observed radio halo emission is assumed to Compton
scatter CMB photons up into the X-ray band. In this specific scenario,
and in the absence of other non-thermal or quasi-thermal X-ray
emission components, the combination of radio and IC X-ray flux
measurements can be used to estimate the magnetic field
strength (\eg Rephaeli 1979; Fusco-Femiano \etal 1999; Blasi \&
Colafrancesco 1999; Eckert \etal 2008).  We have carried out such an
analysis for the Bullet Cluster, 1E\,0657-56 assuming that the
observed power-law X-ray components are entirely due to IC
emission. In this case,

\begin{equation} \frac{F_X}{F_R} =
\frac{E_{ph}}{B^2/8\pi}\left(\frac{\gamma_X}{\gamma_R}\right)^2
\frac{N\left(\gamma_X\right)}{N\left(\gamma_R\right)} =
\frac{E_{CMB}}{B^2/8\pi}\left(\frac{\gamma_X}{\gamma_R}\right)^{1-2\alpha},
\end{equation}

\noindent where $N(\gamma)$ is the number density of electrons with
Lorentz factor $\gamma$, $\gamma_X = (\nu_X/\nu_{CMB})^{1/2}$,
$\gamma_R = (\nu_R/\nu_{cyc})^{1/2}$, $\nu_X$ is the frequency of the
X-rays, $\nu_{CMB}$ is the frequency of the up-scattered CMB photons,
$\nu_R$ is the frequency of the measured radio flux, and $\nu_{cyc}$
is the cyclotron frequency. $E_{CMB}$ is the energy density of CMB
photons at the cluster and $\alpha=\Gamma-1$ is the spectral index of
the power-law components.

Within the context of this specific model, we
infer a peak B field of ~0.15 $\mu$G associated with the regions of 
strongest non-thermal-like X-ray emission.
Here we have used the photon index for each region as
measured from the X-ray data. If we had instead assumed a fixed photon
index $\Gamma=2.0$, then the implied B fields would have been several
times higher.

Two factors argue against this simplest IC interpretation. 
Firstly, the photon indices of the power-law X-ray components
are flatter than typical radio halo synchrotron spectra ($2.0\approxlt
\Gamma \approxlt 2.3$). This implies, within the context of the IC
model, that the same population of electrons are $not$ responsible for
both the radio and X-ray emission.  Secondly, such low magnetic field
values and spatial field configurations appear somewhat at odds with Faraday
rotation studies that measure fields of a few, to a few tens, of
$\mu$G in clusters, with the strongest fields observed in cluster
cores (\eg Kim \etal 1991; Feretti \etal 1995; Taylor \etal 1999,
2001, 2002, 2006, Clarke, Kronberg \& B\"ohringer 2001; Allen \etal 2001;
Govoni \& Feretti 2004).

Theoretically, the spectrum of electrons with Lorentz factor $\gamma
\sim 1000$ responsible for IC X-ray emission can be flatter than those
responsible for the GHz radio emission. Indeed, this appears to be a
general prediction of more detailed calculations of the coupled,
time-dependent properties of electrons, hadrons and plasma waves in a
turbulent intracluster medium, such as will be generated following a
major merger event (\eg Brunetti \etal 2004; Brunetti \& Blasi 2005;
Petrosian \& Bykov 2008). A number of authors have also argued that
the discrepancies between Faraday rotation and IC estimates of the
magnetic field may be reconciled (or at least reduced) by employing
more realistic distributions for the non-thermal particles and
magnetic fields (\eg Petrosian 2001; Carilli \& Taylor 2002; Newman,
Newman \& Rephaeli 2002; Kuo \etal 2003; Brunetti \etal 2004; Govoni
\& Feretti 2004). Thus, although the simplest IC scenario described by
equation 1 is challenged by our results, more sophisticated treatments
may prove consistent with the data.

A second, alternative
interpretation for the observed power-law X-ray components is
bremsstrahlung emission from supra-thermal electrons, energized by
collisions with an energetic non-thermal proton population in the
clusters (Liang \etal 2002; Dogiel \etal 2007; Wolfe \& Melia 2008;
see also Ensslin \etal 1999, Sarazin \& Kempner 2000, Blasi 2000).
The observed spectra appear consistent with this model, although
physical challenges remain (\eg Petrosian \& Bykov 2008; Petrosian \&
East 2008).  

Thirdly, it remains possible that a contribution to the
observed non-thermal-like X-ray flux could arise from synchrotron emission
from ultra-relativistic electrons and positrons, which would also be
characterized by hard power-law spectra ($\Gamma \sim 1.5$; Inoue
\etal 2005). In this case, however, one might expect the hard X-ray
emission to be brighter in the outskirts of the clusters, where the
accretion shocks are strongest.

\subsection{The potential impact of temperature and metallicity
substructure.}

A fourth, possible explanation for the results on non-thermal-like,
power-law signatures is the presence of significant temperature and/or
metallicity structure in the clusters.

In order to explore the potential impact of such effects, we have generated
and analyzed simulated X-ray data sets with characteristics similar to
the Chandra observations of Abell 2319 (similar instrumental response,
exposure time, flux, redshift, mean temperature and absorbing Galactic
column density).  As with the real data, the simulated data sets each have
ten spectral regions with $10^4$ counts. We incorporate a broad
range of temperature and metallicity distributions in and across the
10 regions.  In each case, the mean emission-weighted temperatures in
the ten spatial regions were constrained to match the observed values
for Abell 2319.  The simulated data sets were fitted with the same
basic models as the real data (models A--D) and the improvements in
the fits, i.e. the $\Delta C/\Delta \nu$ values, obtained for models B,C and D
with respect to model A were noted.

We first examined whether the observed power-law signatures could be
accounted for by hot, shocked gas with $kT\sim20$keV.  The flux in the
20keV component was allowed to account for up to 50 per cent of the
total in each region. In all cases, the $\Delta C/\Delta \nu$ values obtained
with this model were much smaller than the observed values. We
conclude that the observed power-law components are unlikely to be due
simply to the presence of unresolved hot, shocked gas in the clusters.

We next examined the potential impact of broad, temperature
distributions, such as might occur if the gas in each spatial region
were highly multiphase. A range of uni-modal temperature models, with
normal, log-normal, skew-normal or T-distributions of varying widths
(widths ranged from 1-5 keV for normal, skew-normal, and T-distributions or
0.1-0.6 for lognormal distributions) were explored.  Although for $individual$
regions broad temperature distributions can sometimes generate
power-law-like signatures, in no case did the combined results for all
10 regions provide comparable $\Delta C/\Delta \nu$ values for models B--D to
those observed.

All of the radio halo clusters appear to have experienced recent,
massive merger events. Such events can can introduce large amounts of
relatively low temperature gas into cluster cores (e.g. the `bullet'
in 1E\,0657-56).  We have therefore examined the potential impact of broad,
bimodal temperature distributions in the clusters. The gas in each
region was assumed to contain two dominant components, with mean
temperatures and luminosities scaled according to standard virial
relations (Mantz et al. 2009, in prep; mass ratios in the range 2:1 to 4:1
were examined). Each component was allowed to have its own broad range
of temperature components, again represented by normal, log-normal,
skew-normal or T-distributions of varying widths.  Although this model
provides larger $\Delta C/\Delta \nu$ values than the above, again, in no
case did the combined results for all 10 regions provide comparable
$\Delta C/\Delta \nu$ improvements to the real data.

We next examined the impact of metallicity variations in the
clusters. Galaxy clusters are commonly observed to contain strong
metallicity gradients, with the strongest gradients observed in
relaxed, cool core clusters, although gradients also observed in some
merging systems (\eg Allen \& Fabian 1998, De Grandi \& Molendi 2001,
Leccardi \& Molendi 2008). When observed in projection, inner
high-metallicity gas will be grouped with outer, lower-metallicity
material, which can complicate the interpretation of spectra.  We have
examined the potential impact of such effects. We assume that each
spectral region contains a bimodal metallicity distribution, with
metallicities differing by factors from 2:1 to 10:1, and relative
fluxes in these components ranging from 1:10 to 10:1. In the first
case, with $isothermal$ gas in each region, the impact of the
metallicity variations is limited.  However, if both metallicity $and$
temperature variations are allowed - be it in a uni-modal or bimodal
manner for the temperature - then significant $\Delta C/\Delta \nu$ values,
comparable to the observations, can be obtained.  The maximum $\Delta
C/\Delta \nu$ values are observed for {\it roughly equal} fluxes in the low and high
metallicity components.  For such cases, we have also confirmed that
allowing the power-law index in each region to vary independently
leads to consistent results on the distributions of photon indices, as
well as flux in the inferred power-law components.  For example, a
model with a 3:1 mass merger, a T-distribution of temperature, 
and a 1:1 mix of high and low metallicity gas satisfies all requirements.

\begin{figure*}
\hbox{
\hspace{0.0cm}
\includegraphics[width=0.51 \textwidth]{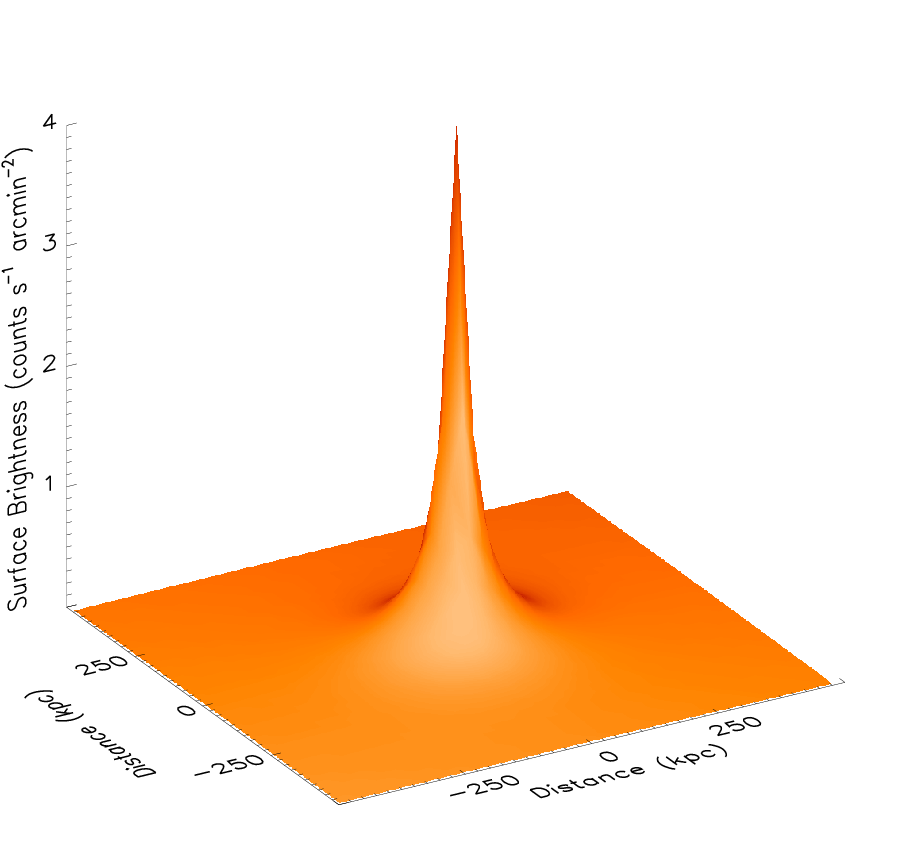}
\hspace{-0.2cm}
\includegraphics[width=0.51 \textwidth]{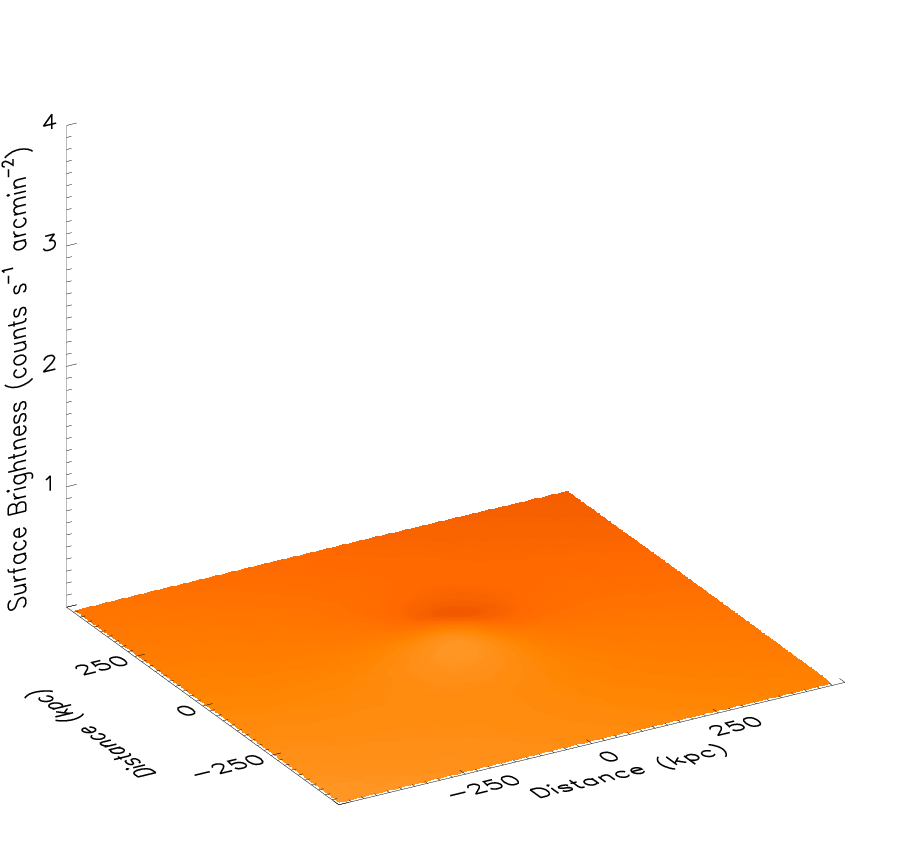}
}
\caption{
3-dimensional representation of the projected surface brightness for
the cool core cluster  
Abell 2029 (left) and the radio halo cluster Abell 2319 (right) scaled
to appear as they would if observed at the same redshift.
The flat surface brightness
core of Abell 2319 with respect to that of Abell
2029 (core radius of 120 kpc versus 20 kpc, respectively)
is the most obvious morphological distinction
and impacts on the relative importance of projection effects 
in the two systems.  
The X and Y axes span 1 Mpc on a side.  The Z axis shows the surface 
brightness in units of counts s$^{-1}$ arcmin$^{-2}$.
}
\label{fig:3d}
\end{figure*}

It is important to consider why, if temperature and metallicity
variations can together explain the observed power-law signatures in
the clusters, are these only detected in radio halos? A possible
resolution lies in the differences in central X-ray surface brightness
of relaxed cool core and merging, radio halo systems: the central
surface brightness is typically an order of magnitude higher in
relaxed cool-core clusters (Figure \ref{fig:3d}). 
For example, cool core clusters might have
beta-model core radii of $0.02$ virial radii or less, whereas for
massive, merging systems a value more like $0.2$ virial radii is
observed (e.g. Allen 1998). For illustration, we consider a step-like
metallicity distribution ranging from 0.5 solar within 0.2 virial
radii to 0.1 solar beyond.  When viewed in projection, a region
spanning the central 100kpc radius for the cool core cluster will have
only $\sim$25 per cent of the flux in the lower metallicity component. In
contrast, for the merging system with the larger core, the flux in the
outer, lower metallicity component will be closer to 50 per cent, for
which maximum $\Delta C/\Delta \nu$ values with models B--D, with respect to
model A, are observed. We note that this may also explain the requirements
for extra spectral
components at radii of $100-200h_{70}^{-1}$ kpc using model D in Abell 1795 and
Abell 2029, since for these annuli also the contribution of the 
projected to the total measured flux approaches 50 per cent. 
(For Abell 2029, however, it also remains possible that the signature 
may in part be generated by processes associated with 
the mini radio halo in this cluster reported by Govoni \etal 2009, 
which also has an extent of approximately $200h_{70}^{-1}$kpc.)

Finally, for completeness, we have also carried out simulations in
which true, non-thermal power-law components were included in the
clusters, with photon indices in the range $1.25<\Gamma<2.5$ and
approximately 10 per cent of the total flux in the power-law component
(as observed).  For photon indices in the range $1.5-2.0$, $\Delta C/\Delta \nu$
values consistent with the observed values are also be obtained.

We conclude that temperature and metallicity variations, similar to
those described above, could be responsible for the observed,
non-thermal-like, power-law components in the clusters. However, these
signatures may also be due to true non-thermal or quasi-thermal
processes, as discussed in Section 5.2.

\section{Summary}

Using high-quality Chandra data, we have
presented new evidence for spatially-extended, non-thermal-like 
X-ray emission components in five of a sample
of seven luminous, merging, radio-halo clusters.  
A control sample of five, comparably X-ray luminous and mostly relaxed
clusters without radio halos do not, in general, 
exhibit similar emission signatures.

The spectra of the detected non-thermal-like
emission components can be
approximated by simple power-law models with photon indices
$1.5<\Gamma<2.0$.  The data for the Bullet
Cluster 1E\,0657-56, demonstrate a spatial correlation between
the regions of highest thermal
pressure, strongest radio halo emission, and strongest power-law emission 
in the cluster core. Our
detailed X-ray maps reveal complex thermodynamic structure in the
radio halo clusters, driven by the ongoing merger activity. 
We confirm the presence of the 
previously-discovered shock front in 1E\,0657-56.
We also report
the discovery of a clear, large-scale shock front in Abell 2219, 
as well as indications of further shock activity in other
clusters (see Appendix). 
 
We have highlighted several possible explanations for the observed,
power-law X-ray emission components.  One possibility is IC
emission; however, the spectra of the electron populations responsible
the IC X-ray emission must be flatter than those producing the radio
halo flux. The magnetic field strengths ($B\sim ~0.15 \mu$G) implied by
the IC model in the regions of the strongest non-thermal-like X-ray
emission are also less than those typically inferred from Faraday
rotation measurements. Both difficulties can, however, be understood
(at least partially) within the context of more sophisticated
treatments of particle acceleration in a turbulent intracluster medium
(\eg Petrosian 2001; Carilli \& Taylor 2002; Newman, Newman \&
Rephaeli 2002; Kuo \etal 2003; Brunetti \etal 2004; Govoni \& Feretti
2004; Petrosian \& Bykov 2008).  A second interpretation in
terms of bremsstrahlung emission from supra-thermal electrons,
energized by collisions with a non-thermal proton population, (Liang
\etal 2002; Dogiel \etal 2007; Wolfe \& Melia 2008; see also Pfrommer
2008) also appears consistent with the observed X-ray spectra, although
physical challenges with this model remain (\eg Petrosian \& Bykov
2008; Petrosian \& East 2008). A contribution to the non-thermal
X-ray flux due to synchrotron emission from ultra-relativistic
electrons and positrons (Inoue \etal 2005) is possible, although
we might expect such a component to be brightest in the cluster
outskirts, where the accretion shocks are strongest. 

Fourthly, and perhaps most interestingly, 
we have shown that power-law signatures, similar to those 
reported here, can also be generated by temperature and/or metallicity 
substructure in the cluster, particularly in the presence of 
metallicity gradients. In this case, an important distinguishing
characteristic between radio halo and non radio halo, cooling-core clusters is 
the more sharply-peaked X-ray emission of the latter population.

Our results have clear implications for the interpretation of 
`soft-excess' X-ray emission from galaxy clusters
(\eg Durret \etal 2008 and references therein) and show that apparent
soft excesses (or best-fit absorption of less than the Galactic value)
can arise from inadequate modelling of the emission spectra in the
presence of significant spatial variations in temperature and metallicity,
as well as additional emission components.

Further X-ray and radio mapping, coupled with new $\gamma$-ray and TeV
observations and detailed hydrodynamical simulations will be vital for
improving our understanding of the origin of non-thermal and
quasi-thermal X-ray emission from galaxy clusters.

Finally, we note that analyses such as ours push at the limits of what
is possible with Chandra. However, our restriction to using only data
gathered above a conservative background-to-total counts threshold,
the null results obtained for the control sample of non-radio halo
clusters, the insensitivity of the results to the precise energy band
used all argue that the basic findings
are likely to be robust.

\section{Acknowledgments}

We are grateful to Haida Liang for kindly providing the radio data for
1E\,0657-56, and Jeremy Sanders for the contour binning algorithm.  We
thank Roger Blandford, Gianfranco Brunetti, Stefan Funk, Haida Liang,
Greg Madejski, Vahe Petrosian, Christoph Pfrommer, Anita Reimer, Olaf
Reimer, Jeremy Sanders, and Norbert Werner for helpful comments.  We
thank R. Glenn Morris for discussions and technical support, and
H. Marshall and A. Vikhlinin for discussions regarding Chandra
calibration. All computational analysis was carried out using the
KIPAC XOC compute cluster at Stanford University and the Stanford
Linear Accelerator Center (SLAC).  We acknowledge support from the
National Aeronautics and Space Administration through Chandra Award
Numbers G06-7123X, issued by the Chandra X-ray Observatory Center,
which is operated by the Smithsonian Astrophysical Observatory for and
on behalf of the National Aeronautics and Space Administration under
contract NAS8-03060. This work was supported in part by the
U.S. Department of Energy under contract number DE-AC02-76SF00515.

\section*{Appendix: detailed thermodynamic mapping of other clusters 
in the sample}

This appendix includes thermodynamic X-ray maps for all other
clusters in the sample (in addition to 1E\,0657-56 and Abell 2029)
that have $\sim$20 or more regions containing 3,000 counts.  Our
results are broadly consistent with, but refine and expand on,
previous studies (\eg Markevitch \& Vikhlinin 2001; Markevitch,
Vikhlinin \& Mazzotta 2001; Fabian \etal 2001; Worrall \& Birkinshaw
2003; Sun \etal 2003; Kempner \& David 2004a; Kempner \& David 2004b;
Boschin \etal 2004; Govoni \etal 2004; O'Hara, Mohr \& Guerrero 2004;
Sanders, Fabian \& Taylor 2005; Sakelliou \& Ponman 2006; Dupke \etal
2007). 

A key result is the discovery of a large-scale shock front 
in Abell 2219 (Fig. 8b).  The cluster 
shows a clear, arc-shaped region of hot, $kT\approxgt16$ keV
(projected) gas roughly 2 arcmin to the northwest of the cluster
center. This gas also has high pressure and entropy and is almost
certainly associated with a merger-driven shock front. Together with
1E\,0657-56 and Abell 520, Abell 2219 is only the third X-ray luminous
cluster in which a clear, large-scale shock front, viewed
approximately edge-on, has been discovered (see Markevitch \&
Vikhlinin 2007 for a review).  For more detail see Applegate \etal 2009,
in prep.

We have estimated the Mach number and velocity of the shock in Abell
2219.  The deprojected pre- and post-shock temperatures of
$13^{+3}_{-2}$ keV and $25^{+11}_{-8}$ keV, respectively, lead to an
estimate for the Mach number of $1.9^{+0.7}_{-0.6}$. A density jump of
a factor $\sim$1.5 is also seen at the front, which leads to an
independent and consistent estimate for the Mach number of
$1.3^{+0.2}_{-0.2}$. Given an ambient sound speed of $\sim 1850$ km/s,
estimated from the pre-shock temperature in the northwest region, we
determine a shock velocity of $\sim2500$ km/s.
Deeper $Chandra$ observations are required to enable a more robust
investigation of the shock properties in Abell 2219.
Interestingly, we note that Abell 2219 also exhibits a region of
exceptionally hot, $kT=22^{+7}_{-4}$ keV, high pressure gas
coincident with the peak of the X-ray surface brightness, which is also,
presumably, associated with recent shock activity in the cluster core.

Abell 2744, like Abell 2219, does not provide significant evidence for 
non-thermal-like X-ray emission.  Its X-ray maps show 
a number of interesting features.  The cluster is in the process of a massive,
merger event.
The overall pressure maximum is
coincident with the X-ray surface brightness peak and is associated
with the main (southeastern) cluster core.  The hottest gas ($kT\sim
11$ keV) is also found in this region.  The complex entropy structure
in the main cluster suggests possible shocks and significant
ongoing turbulence possibly indicating that the colliding cores have 
already passed through each other. 
The coolest ($kT\sim 6$ keV), lowest entropy gas
in Abell 2744 resides in the core of the merging subcluster, to the
northwest of the main cluster.

For Abell 2319 and 2163, the thermodynamic maps are complex and
confirm that the clusters have experienced massive, recent merger
events. For both systems, the temperature, pressure and entropy maps
show a high degree of asymmetry. As with 1E\,0657-56, we observe a
close correlation between regions of strongest apparent dynamical
activity and brightest radio halo emission in the clusters (see Govoni
\etal 2004 for the radio images).  In the case of Abell 2319, we observe
a clear extension of cool, low entropy gas to the northwest; the radio
halo emission is also extended in this direction. For Abell 2163, the
coolest, lowest-entropy gas is located to the west of the main X-ray
brightness peak. As with the Bullet Cluster, this gas may represent
the surviving core of a recently merged subcluster.  
Previous claims of
hard, non-thermal X-ray emission from Abell 2319 and 2163, based on
RXTE data, were reported by Gruber \& Rephaeli (2002) and Rephaeli,
Gruber \& Arieli (2006), respectively.  (An upper limit for Abell 2163
was also reported by Feretti \etal (2001), using BeppoSax.) These
results, which probe harder X-ray energies but with limited spatial
resolution, are broadly consistent with ours.

For Abell 478 and 1795, the thermodynamic maps indicate relatively
relaxed thermodynamic states with a high degree of symmetry. The peaks
of the thermal X-ray brightness coincide with the locations of the
dominant cluster galaxies.  The X-ray peaks mark the locations of the
lowest temperature and entropy, and highest thermal
pressure. Unsurprisingly, and as with Abell 2029, neither cluster is
in $perfect$ equilibrium, as evidenced by detailed structure observed
in the X-ray image, especially at small radii (see \eg Fabian \etal
2001; Markevitch, Vikhlinin \& Mazzotta 2001; Sun \etal 2003).  In
particular, for Abell 1795, some approximate north-south elongation 
is visible in the pressure
and entropy maps that may be related to recent dynamical activity and 
the observed peculiar motion
of the dominant cluster galaxy (Fabian \etal 2001).

\begin{figure*}
\hbox{
\hspace{0.0cm}
\includegraphics[width=0.47 \textwidth]{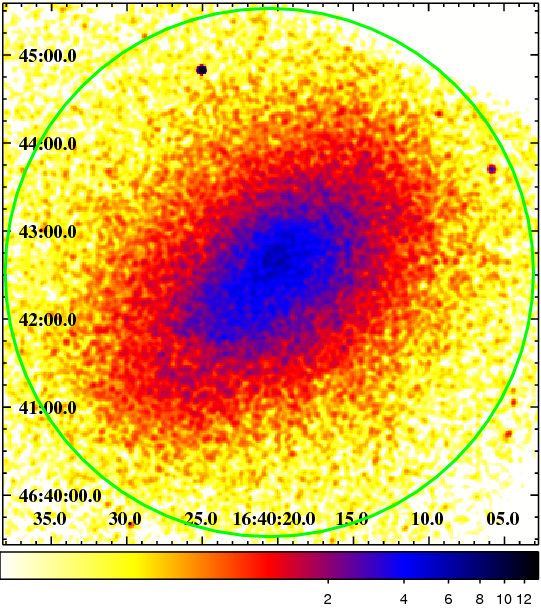}
\hspace{0.9cm}
\includegraphics[width=0.47 \textwidth]{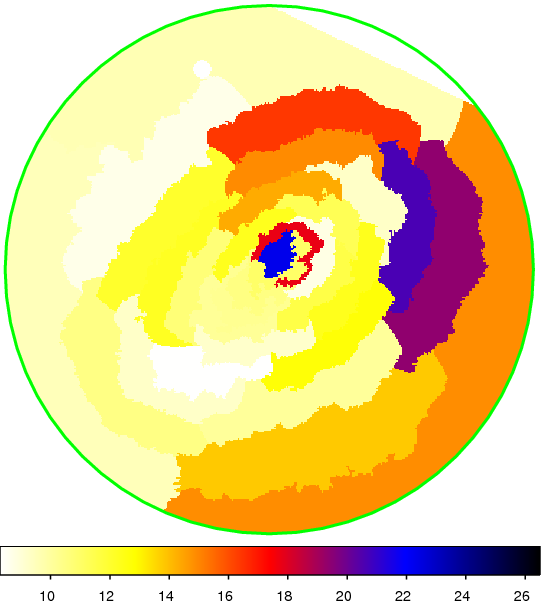}
}
\hbox{
\hspace{0.0cm}
\includegraphics[width=0.47 \textwidth]{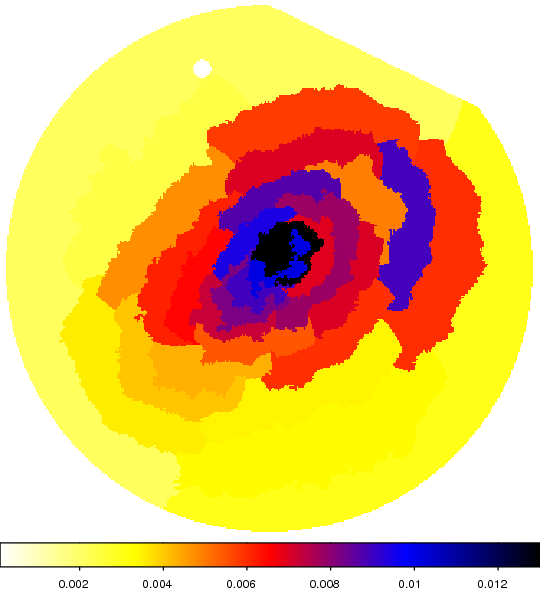}
\hspace{0.9cm}
\includegraphics[width=0.47 \textwidth]{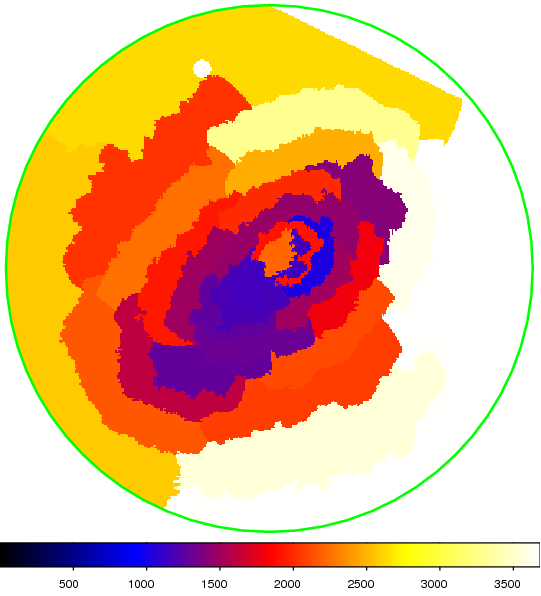}
}
\caption{
Thermodynamic maps for Abell 2219. Note the clear shock front
approximately 2 arcmin to the northwest of the cluster center. The
density and temperature jumps at the shock front indicate a Mach
number of 1.3-1.9 (see also Applegate \etal 2009, in prep). 
Note also the very hot gas ($kT = 22^{+7}_{-4}$ keV) coincident with
the X-ray surface brightness peak which may also be associated with
recent/ongoing shock activity. Other details as for Fig.
\ref{fig:bulletT}.
}
\label{fig:a2219}
\end{figure*}

\begin{figure*}
\hbox{
\hspace{0.0cm}
\includegraphics[width=0.47 \textwidth]{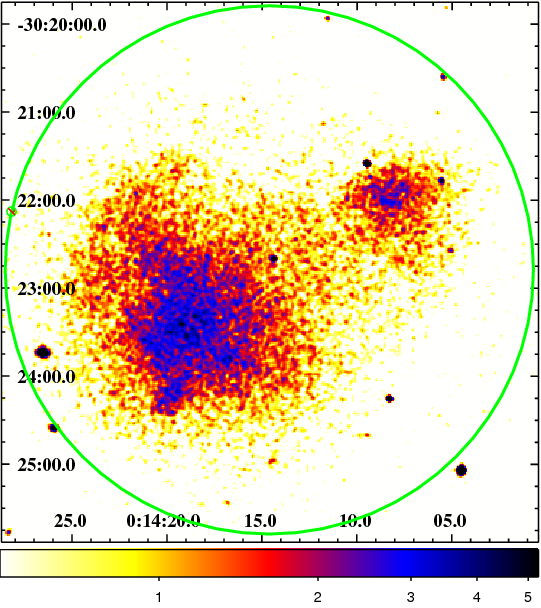}
\hspace{0.9cm}
\includegraphics[width=0.47 \textwidth]{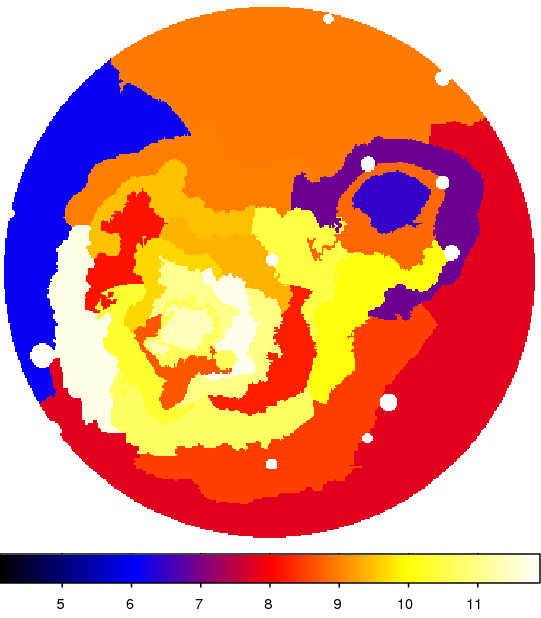}
}
\hbox{
\hspace{0.0cm}
\includegraphics[width=0.47 \textwidth]{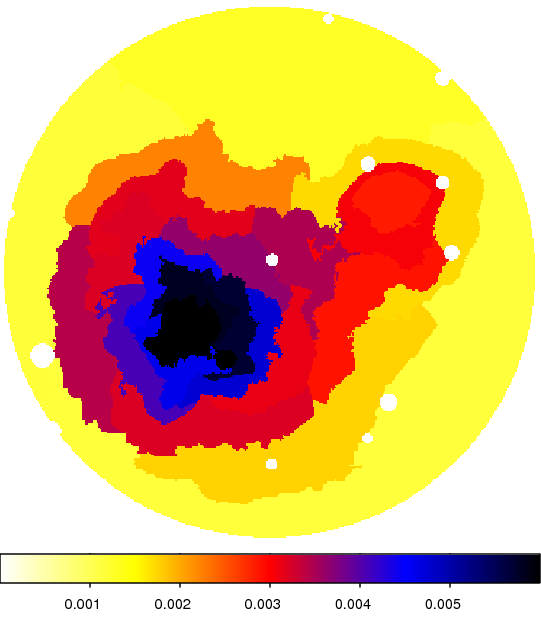}
\hspace{0.9cm}
\includegraphics[width=0.47 \textwidth]{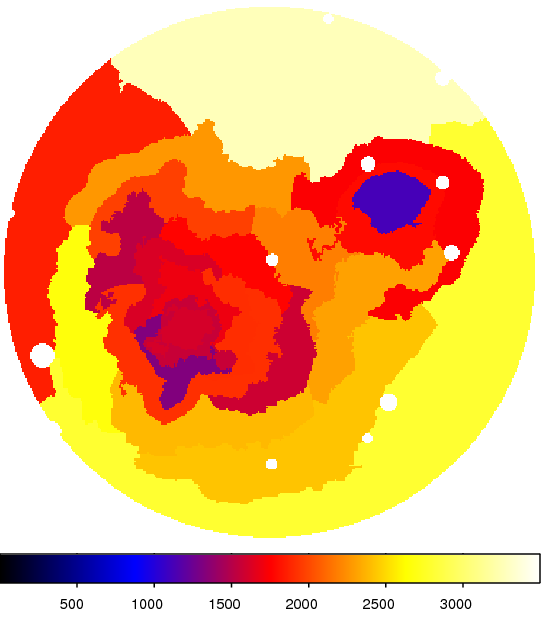}
}
\caption{
Thermodynamic maps for Abell 2744. 
The cluster is in the process of a major, ongoing merger event.
The complex entropy structure in
the main cluster core suggests significant dynamical activity.  The
coolest, lowest entropy gas is found in the core of the merging
subcluster, to the northwest. Other details as for Fig.
\ref{fig:bulletT}.  }
\label{fig:a2744}
\end{figure*}

\begin{figure*}
\hbox{
\hspace{0.0cm}
\includegraphics[width=0.47 \textwidth]{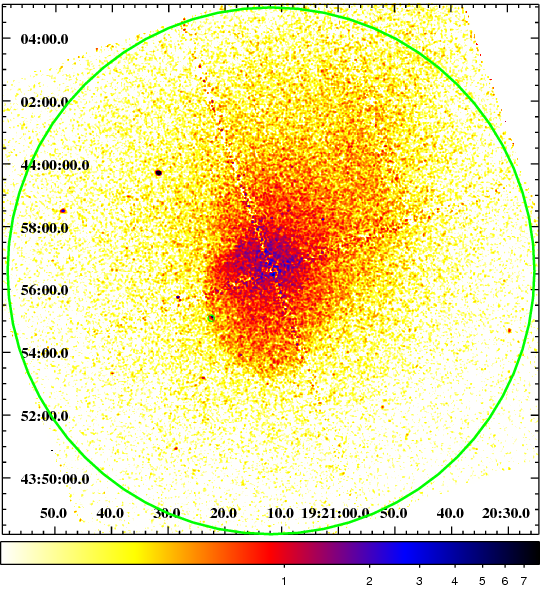}
\hspace{0.9cm}
\includegraphics[width=0.47 \textwidth]{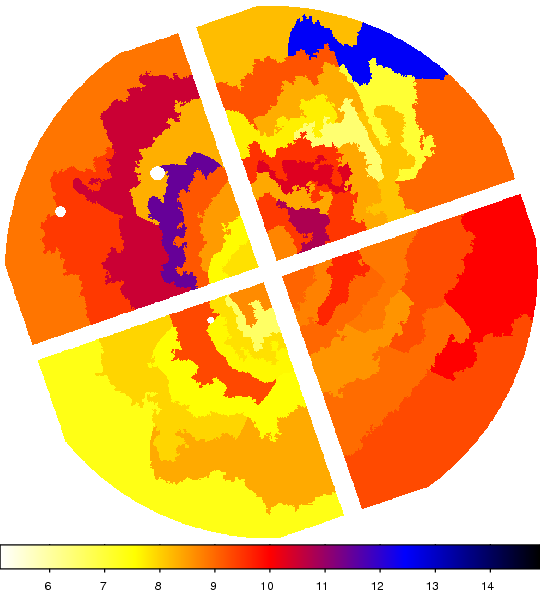}
}
\hbox{
\hspace{0.0cm}
\includegraphics[width=0.47 \textwidth]{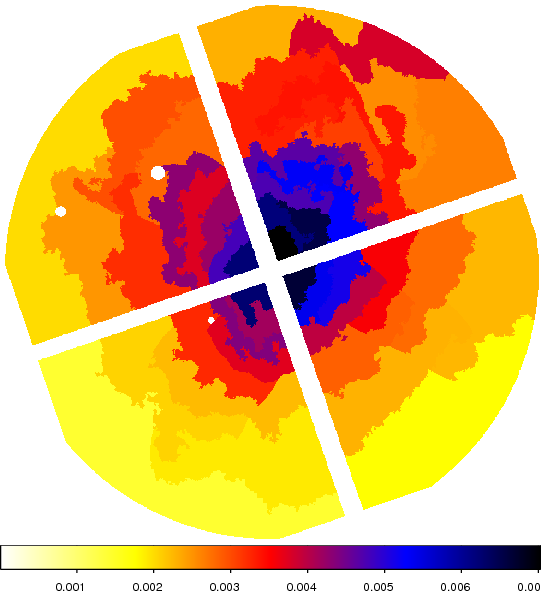}
\hspace{0.9cm}
\includegraphics[width=0.47 \textwidth]{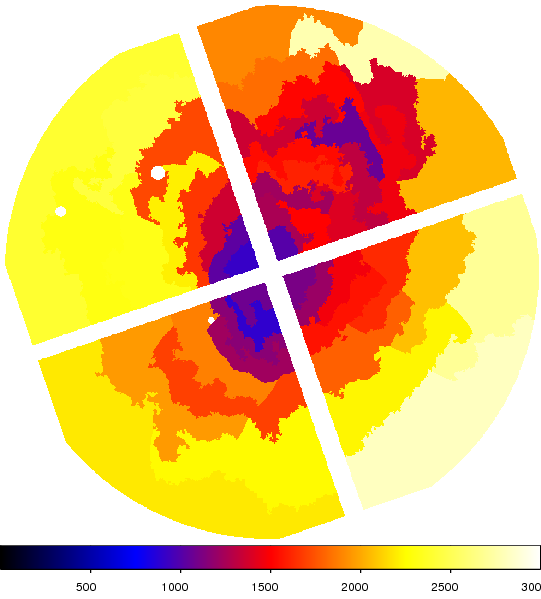}
}
\caption{
Thermodynamic maps for Abell 2319. The maps reveal the complex
dynamical state of the cluster. We observe a clear extension of cool,
low entropy gas to the northwest. Other details as for Fig.
\ref{fig:bulletT}.
}
\label{fig:a2319}
\end{figure*}

\begin{figure*}
\hbox{
\hspace{0.0cm}
\includegraphics[width=0.47 \textwidth]{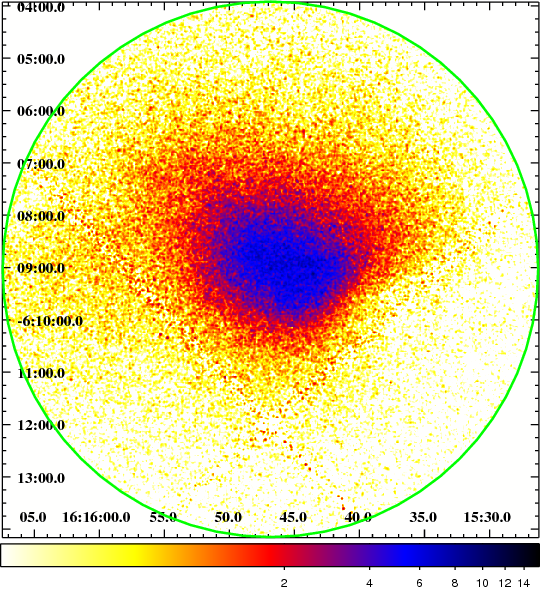}
\hspace{0.9cm}
\includegraphics[width=0.47 \textwidth]{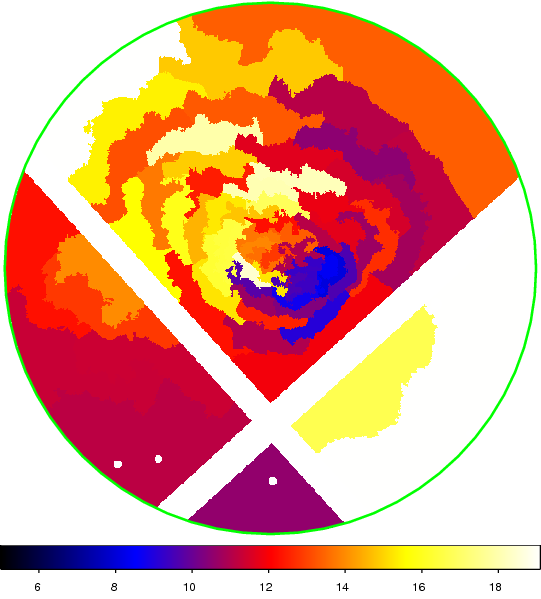}
}
\hbox{
\hspace{0.0cm}
\includegraphics[width=0.47 \textwidth]{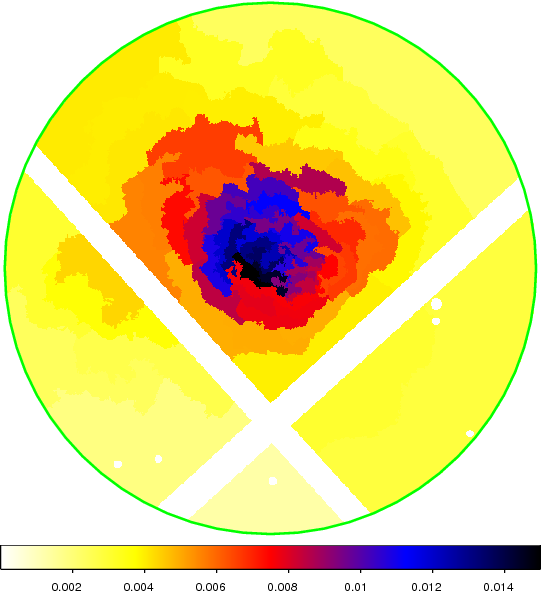}
\hspace{0.9cm}
\includegraphics[width=0.47 \textwidth]{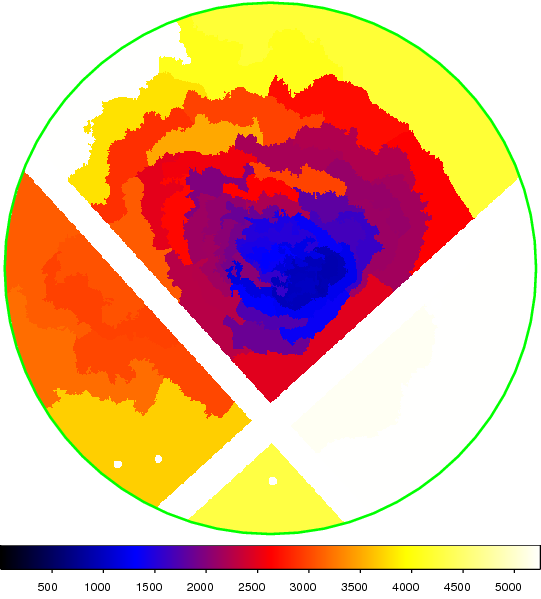}
}
\caption{
Thermodynamic maps for Abell 2163. The maps reveal the complex
dynamical state of the cluster. The coolest, lowest-entropy
gas is located to the west of the main X-ray brightness peak. As with
the Bullet Cluster, this gas may represent the surviving core of a
recently merged subcluster.  The maps also indicate a possible shock
front $\sim1.3$ arcmin to the north of the main surface brightness
peak. Other details as for Fig. \ref{fig:bulletT}.
}
\label{fig:a2163}
\end{figure*}

\begin{figure*}
\hbox{
\hspace{0.0cm}
\includegraphics[width=0.47 \textwidth]{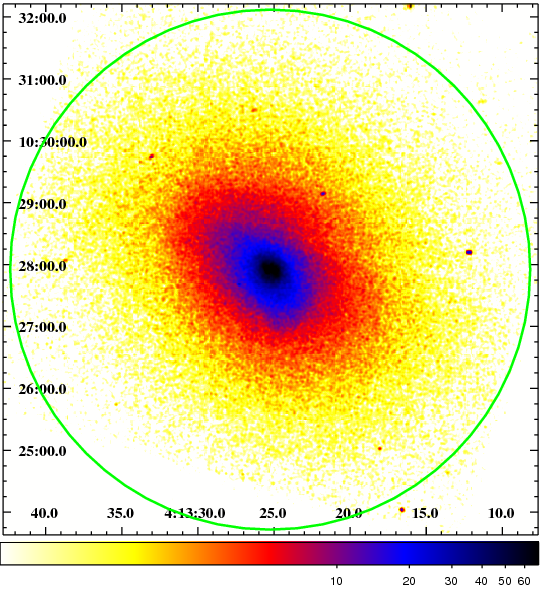}
\hspace{0.9cm}
\includegraphics[width=0.47 \textwidth]{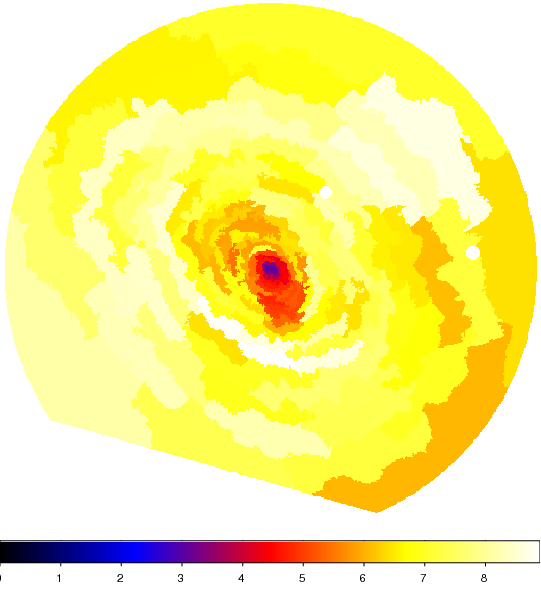}
}
\hbox{
\hspace{0.0cm}
\includegraphics[width=0.47 \textwidth]{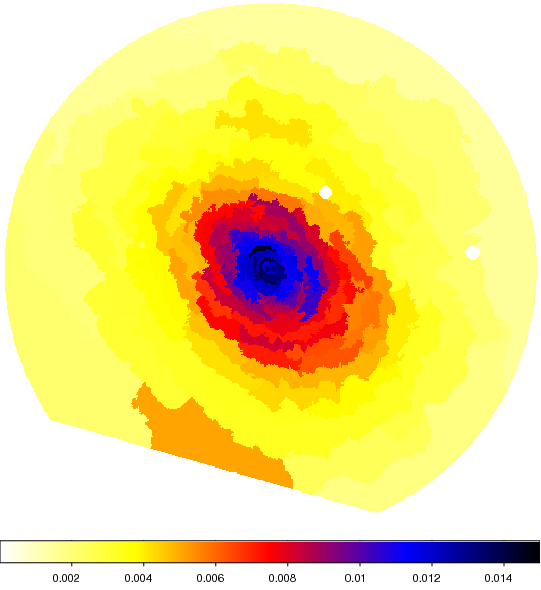}
\hspace{0.9cm}
\includegraphics[width=0.47 \textwidth]{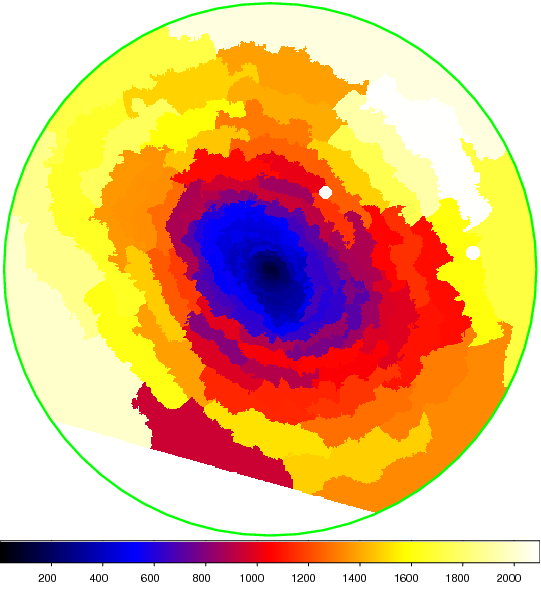}
}
\caption{
Thermodynamic maps for Abell 478. The cluster exhibits a relaxed dynamical
state. Other details as for Fig. \ref{fig:bulletT}.
}
\label{fig:a478}
\end{figure*}

\begin{figure*}
\hbox{
\hspace{0.0cm}
\includegraphics[width=0.47 \textwidth]{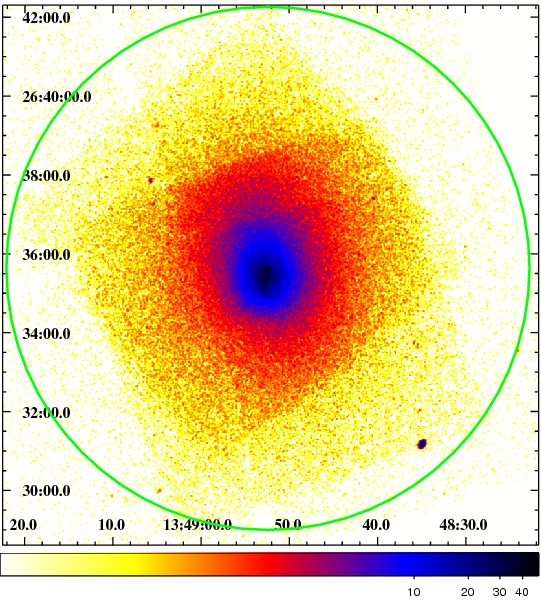}
\hspace{0.9cm}
\includegraphics[width=0.47 \textwidth]{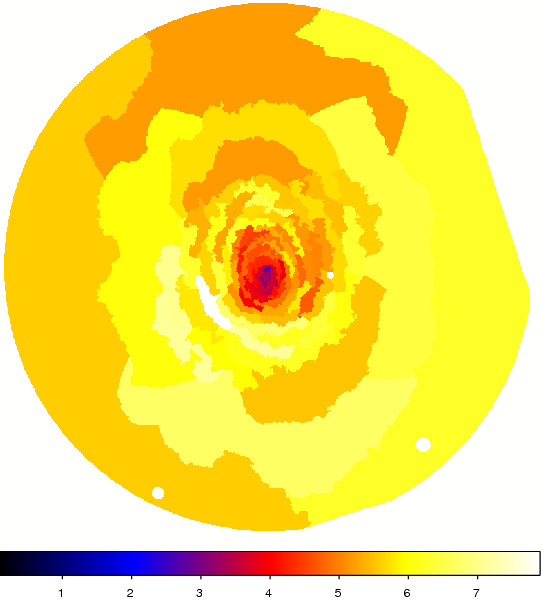}
}
\hbox{
\hspace{0.0cm}
\includegraphics[width=0.47 \textwidth]{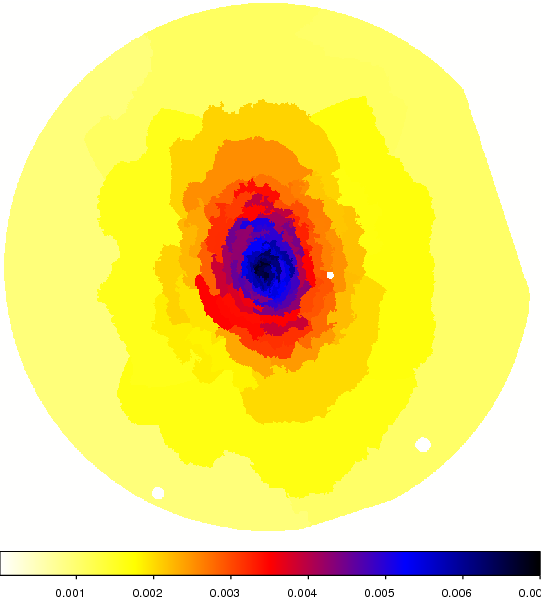}
\hspace{0.9cm}
\includegraphics[width=0.47 \textwidth]{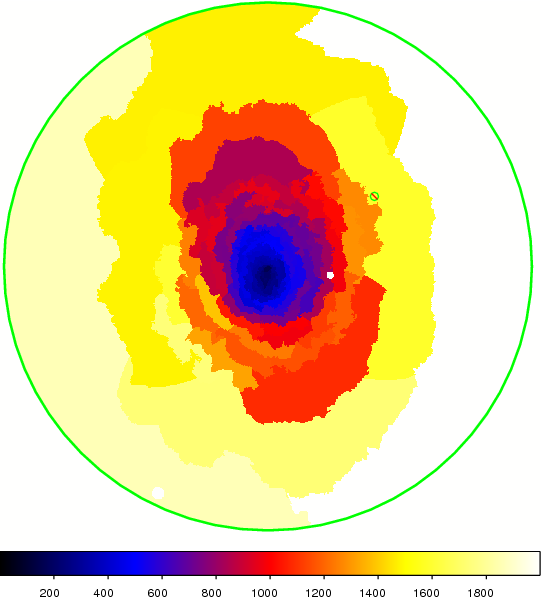}
}
\caption{
Thermodynamic maps for Abell 1795. The cluster exhibits a generally 
relaxed dynamical state, although some asymmetry is visible in the 
pressure and entropy maps
that may be related to recent dynamical activity and the observed
peculiar motion of the dominant cluster galaxy (Fabian \etal
2001). Other details as for Fig. \ref{fig:bulletT}.
}
\label{fig:a1795}
\end{figure*}


\begin{thebibliography}{}
\bibitem{} Ajello M., Rebusco P., Cappelluti N., Reimer O., B\"ohringer H., Greiner J., Gehrels N., Tueller J., Moretti A., 2009, ApJ, 690, 367
\bibitem{} Allen S.W., Fabian A.C., Johnstone R.M., White D.A., Daines S.J., Edge A.C., Stewart G.C., 1993, MNRAS, 262, 901
\bibitem{} Allen S.W., Fabian A.C., 1998, MNRAS, 297L, 57
\bibitem{} Allen S.W., 1998, MNRAS, 296, 392
\bibitem{} Allen S.W., Schmidt R.W., Fabian A.C., 2001, MNRAS, 328, 27
\bibitem{} Allen S.W., Taylor G.B., Nulsen P.E.J., Johnstone R.M., David L.P.,
Ettori S., Fabian A.C., Forman W., Jones C., McNamara B., 2001, MNRAS, 324, 842
\bibitem{} Allen S.W., Schmidt R.W., Fabian A.C., 2002, MNRAS, 335, 256
\bibitem{} Allen S.W., Rapetti D.A., Schmidt R.W., Ebeling H., Morris R.G., Fabian A.C., 2008, MNRAS, 383, 879
\bibitem{} Arabadjis J., Bregman J.N., 2000, ApJ, 536, 144
\bibitem{} Arnaud, K.A., 1996, in Astronomical Data Analysis Software and Systems V, eds. Jacoby G. and Barnes J., ASP Conf. Series volume 101, p17
\bibitem{} Balucinska-Church M., McCammon D., 1992, ApJ, 400, 699
\bibitem{} Bevington P.R., 1969, Data Reduction and Error Analysis for the Physical Sciences (New York: McGraw-Hill)
\bibitem{} Blasi P., Colafrancesco S., 1999, APh, 12, 169
\bibitem{} Blasi P., 2000, ApJ, 532L, 9
\bibitem{} Bonamente M., Lieu R., Mittaz J.P.D., 2001, ApJ, 546, 805
\bibitem{} Bonamente M., Joy M., Lieu R., 2003, ApJ, 585, 722
\bibitem{} Boschin W., Girardi M., Barrena R., Biviano A., Feretti L., Ramella M., 2004, A\&A, 416, 839
\bibitem{} Bowyer S., Bergh\"ofer T., Korpela E., 1999, ApJ, 526, 592
\bibitem{} Bowyer S., Korpela E., Lampton M., Jones T.W., 2004, ApJ, 605, 168
\bibitem{} Bregman J.N., Lloyed-Davies E.J., 2006, ApJ, 644, 167
\bibitem{} Brunetti G., Setti G., Feretti L., Giovannini G., 2001, MNRAS, 320, 365
\bibitem{} Brunetti G., Blasi P., Cassano R., Gabici S., 2004, MNRAS, 350, 1174
\bibitem{} Brunetti G., Blasi P., 2005, MNRAS, 363, 1173
\bibitem{} Brunetti G., Lazarian A., 2007, MNRAS, 378, 245
\bibitem{} Buote D.A,. Tsai J.C., 1996, ApJ, 458, 27
\bibitem{} Buote D.A., 2001, ApJ, 553, L15
\bibitem{} Bykov A.M., Bloemen H., Uvarov Y.A., 2000, A\&A, 362, 886
\bibitem{} Carilli C.L., Taylor G.B., 2002, ARA\&A, 40, 319
\bibitem{} Clarke T.E., Kronberg P.P., B\"ohringer H., 2001, ApJ, 547, 111
\bibitem{} Clarke T.E., Blanton E.L., Sarazin C.L., 2004, ApJ, 616, 178
\bibitem{} Cassano R., Brunetti G., Setti G., 2006, MNRAS, 369, 1577
\bibitem{} Cassano R., Brunetti G., Setti G., Govoni F., Dolag K., 2007, MNRAS, 378, 1565
\bibitem{} De Grandi S., Molendi S., 2001, ApJ, 551, 153
\bibitem{} Dickey J.M., Lockman F.J., 1990. ARA\&A, 28, 215
\bibitem{} Dogiel V.A., Colafrancesco S., Ko C.M., Kuo P.H., Hwang C.Y., Ip W.H., Birkinshaw M., Prokhorov D.A., 2007, A\&A, 461, 433
\bibitem{} Dupke R.A., Mirabal N., Bregman J.N., Evrard A.E., 2007, ApJ, 668, 781
\bibitem{} Durret F., Kaastra J.S., Nevalainen J., Ohashi T., Werner N., 2008, SSRv, 134, 51 
\bibitem{} Eckert D., Produit N., Paltani S., Neronov A., Courvoisier T.J.L.,
2008, A\&A, 479, 27
\bibitem{} Elbaz D., Arnaud M., Boehringer H., 1995, A\&A, 303, 345E
\bibitem{} Ensslin T.A., Biermann P.L., Klein U., Kohle S., 1998, A\&A, 378, 777
\bibitem{} Ensslin T.A., Lieu R., Biermann P.L., 1999, A\&A, 344, 409
\bibitem{} Ensslin T.A., Br\"uggen M., 2002, MNRAS, 331, 1011
\bibitem{} Fabian A.C., 1996, Sci, 271, 1244
\bibitem{} Fabian A.C., Sanders J.S., Ettori S., Taylor G.B., Allen S.W., 
Crawford C.S., Iwasawa K., Johnstone R.M., 2001, MNRAS, 321L, 33
\bibitem{} Fabian A.C., Sanders J.S., Taylor G.B., Allen S.W., Crawford C.S., 
Johnstone R.M., Iwasawa K., 2006, MNRAS, 366, 417
\bibitem{} Feretti L., Dallacasa D., Giovannini G., Tagliani A., 1995, A\&A, 302, 680
\bibitem{} Feretti L., Fusco-Femiano R., Giovannini G., Govoni F., 2001, A\&A, 
373, 106
\bibitem{} Feretti L., 2005, Adv. Space Res., 36, 729
\bibitem{} Feretti L., Schuecker P., B\"oringer H., Govoni F., Giovannini G., 2005, A\&A, 444, 157
\bibitem{} Feretti L., Giovannini G., 2007, arxiv:astro-ph/0703494
\bibitem{} Finoguenov A., Briel U., Henry J., 2003, A\&A, 410, 777
\bibitem{} Fujita Y., Hayashida K., Nagai M., Inoue S., Matsumoto H., Okabe N., Reiprich T.H., Sarazin C.L., Takizawa M., 2008, PASJ, 60, 1133
\bibitem{} Fusco-Femiano R., dal Fiume D., Feretti L., Giovannini G., Grandi P., Matt G., Molendi S., Santangelo A., 1999, ApJ, 513L, 21
\bibitem{} Fusco-Femiano R., Orlandini M., Brunetti G., Feretti L,Giovannini G., Grandi P., Setti G., 2004, ApJ, 602L, 73
\bibitem{} Fusco-Femiano R., Landi R., Orlandini M., 2007, ApJ, 654L, 9
\bibitem{} Giovannini G., Feretti L., 2002, in `Merging Processes of Galaxy 
Clusters', eds. Feretti L., Gioia I.M., Giovannini G., ASSL, Kluwer, p. 197
\bibitem{} Gitti M., Ferrari C., Domainko W., Feretti L., Schindler S., 
2007, A\&A, 470, 25
\bibitem{} Govoni F., Feretti L., Giovannini G., B\"ohringer H., Reiprich T.H.,
Murgia M., 2001, A\&A, 376, 803
\bibitem{} Govoni F., Feretti L., 2004, IJMPD, 13, 1549
\bibitem{} Govoni F., Markevitch M., Vikhlinin A., VanSpeybroeck L., Feretti L., Giovannini G., 2004, ApJ, 605, 695
\bibitem{} Govoni F., Murgia M., Markevitch M., Feretti L., Giovannini G., Taylor G.B., Carretti E., 2009/ arXiv:astro-ph/0901.1941
\bibitem{} Gruber D., Rephaeli Y., 2002, ApJ, 565, 877
\bibitem{} Inoue S., Aharonian F.A., Sugiyama N., 2005, ApJ, 628, 9
\bibitem{} Johnstone R.M., Fabian A.C., Edge A.C., Thomas P.A., 1992, MNRAS, 255, 431
\bibitem{} Kaastra J.S., Mewe R., 1993, Legacy, 3, HEASARC, NASA 
\bibitem{} Kalberla P.M., Burton W.B., Hartmann Dap, Arnal E.M., Bajaja E., Morras R., Poeppel W.G.L., 2005, A\&A, 440, 775
\bibitem{} Kempner J.C., David L.P., 2004a, ApJ, 607, 200 
\bibitem{} Kempner J.C., David L.P., 2004b, MNRAS, 349, 385
\bibitem{} Kim K.T., Kronberg P.P., Tribble P.C., 1991, ApJ, 379, 80
\bibitem{} Komatsu E. \etal 2001, PASJ, 53, 57
\bibitem{} Kuo P.H., Hwang C.Y., Ip W.H., 2003, ApJ, 594, 732
\bibitem{} Leccardi A., Molendi S., 2008, A\&A, 487, 461
\bibitem{} Liang H., Hunstead, R.W., Birkinshaw M., \& Andreani P., 2000, ApJ, 544, 686
\bibitem{} Liang H., Dogiel V.A., Birkinshaw M., 2002, MNRAS, 337, 567
\bibitem{} Lieu R., Mittaz J. P. D., Bowyer S., Breen J. O., Lockman F. J., Murphy E. M., Hwang C.-Y., 1996, Sci, 274, 1335
\bibitem{} Markevitch M., 1998, ApJ, 504, 27
\bibitem{} Markevitch M., Vikhlinin A., Mazzotta P., 2001, ApJ, 562L, 153
\bibitem{} Markevitch M., Vikhlinin A., 2001, ApJ, 563, 95
\bibitem{} Markevitch M., Gonzalez A.H., David L., Vikhlinin A., Murray S., Forman W., Jones C., Tucker W., 2002, ApJ, 567, L27
\bibitem{} Markevitch M., 2006, ESA SP-604: The X-ray Universe 2005, 723
\bibitem{} Markevitch M., Vikhlinin A., 2007, PhR, 443, 1
\bibitem{} Murgia M., Govoni F., Markevitch M., Feretti L., Giovannini G., 
Taylor G.B., Carretti E., 2009, arXiv:astro-ph/0901.1943
\bibitem{} Newman W.I., Newman A.L., Rephaeli Y., 2002, ApJ, 575, 755
\bibitem{} O'Hara T.B., Mohr J.J., Guerrero M.A., 2004, ApJ, 604, 604
\bibitem{} Petrosian V., 2001, ApJ, 557, 560
\bibitem{} Petrosian V., 2003, in Bowyer S., Hwang C.-Y., eds, ASP Conf. Ser. Vol. 301, Matter and Energy in Clusters of Galaxies. Astron. Soc. Pac.,
San Francisco, p. 337
\bibitem{} Petrosian V., Madejski, G., Luli K., 2006, ApJ, 652, 948
\bibitem{} Petrosian V., Bykov A.M., SSR, 124, 191
\bibitem{} Petrosian V., East W.E., 2008, arxiv:astro-ph/0802.0900
\bibitem{} Pfrommer C., 2008, MNRAS, 385, 1242
\bibitem{} Pfrommer C., Ensslin T.A., Springel V., 2008, MNRAS, 385, 1211
\bibitem{} Pointecouteau E., Arnaud M., Kaastra J., de Plaa J., 2004, A\&A, 423, 33
\bibitem{} Renaud M., B\'elanger G., Paul J., Lebrun F., Terrier R., 2006, A\&A, 453L, 5
\bibitem{} Rephaeli Y., 1979, ApJ, 227, 364
\bibitem{} Rephaeli Y., Gruber D., Blanco P., 1999, ApJ, 511L, 21
\bibitem{} Rephaeli Y., Gruber D., 2002, ApJ, 579, 587
\bibitem{} Rephaeli Y., Gruber D., Arieli Y., 2006, ApJ, 649, 673
\bibitem{} Rossetti M., Molendi S, 2004, A\&A, 414L, 41
\bibitem{} Sakelliou I., Ponman T.J., 2006, MNRAS, 367, 1409
\bibitem{} Sanders J.S., Fabian A.C., Taylor G.B., 2005, MNRAS, 356, 1022
\bibitem{} Sanders J.S., Fabian A.C., Dunn R.J.H., 2005, MNRAS, 360, 133
\bibitem{} Sanders J.S., 2006, MNRAS, 371, 829
\bibitem{} Sanders J.S., Fabian A.C., 2007, MNRAS, 381, 1381
\bibitem{} Sarazin C.L., Kempner J.C., 2000, ApJ, 533, 73
\bibitem{} Schuecker P. B\"ohringer H., Reiprich T.H., Feretti L., 2001, A\&A, 378, 408 
\bibitem{} Sun M., Jones C., Murray S.S., Allen S.W., Fabian A.C., Edge A.C., 
2003, ApJ, 587, 619
\bibitem{} Taylor G.B., Allen, S.W., Fabian A.C. 1999, in Diffuse Thermal and
Relativistic Plasma in Galaxy Clusters, ed. H. Bo¨hringer, L. Feretti,
P. Schuecker (MPE Rep. 271; Garching: MPE), 77
\bibitem{} Taylor G.B., Govoni F., Allen S.W., Fabian A.C., 2001, MNRAS, 326, 2
\bibitem{} Taylor G.B., Fabian A.C., Allen S.W., 2002, MNRAS, 334, 769
\bibitem{} Taylor G.B., Gugliucci N.E., Fabian A.C., Sanders J.S., Gentile G., Allen S.W., 2006, MNRAS, 368, 1500
\bibitem{} Timokhin A.N., Aharonian F.A., Neronov A.Y., 2004, A\%A, 417, 391
\bibitem{} Vikhlinin A., Markevitch M., Murray S.S., Jones C., Forman W., 
	Van Speybroeck L, 2005, ApJ, 628, 655
\bibitem{} Wik D.R., Sarazin C.L., Finoguenov A., Matsushita K., Nakazawa K.,
Clarke T.E., 2009, accepted in ApJ, astro-ph/0902.3658
\bibitem{} Wolfe B, Melia F., 2008, ApJ, 675, 156
\bibitem{} Worrall D.M., Birkinshaw M., 2003, MNRAS, 340, 1261
\end{thebibliography}
\end{document}